\documentclass[10pt,a4paper]{article}
\usepackage[utf8]{inputenc}
\usepackage{amsmath}
\usepackage{amsthm}
\usepackage{amsfonts}
\usepackage{amssymb}
\usepackage{graphicx}
\usepackage{color}
\usepackage{geometry}
\usepackage{lipsum}
\usepackage[final]{changes}

\usepackage{natbib}
\newtheorem{defi}{Definition}[]
\newtheorem{theorem}{Theorem}[section]
\newtheorem*{proof*}{Proof}
\newtheorem{example}{Example}[section]
\newtheorem{corol}{Corollary}[]
\newtheorem{remark}{Remark}[]
\newtheorem{lemma}{Lemma}[]
\definechangesauthor[color=red]{Fe}

\begin{document}

\title{Context-specific independencies for ordinal variables in chain regression models}
\author{Federica Nicolussi\footnote{Department of Statistics and quantitative methods, 
		University of Milan-Bicocca. Via Bicocca degli Arcimboldi, 8, 20126, Milan, Italy.\textsl{email:} federica.nicolussi@unimib.it}, Manuela Cazzaro\footnote{Department of Statistics and quantitative methods, 
		University of Milan-Bicocca. Via Bicocca degli Arcimboldi, 8, 20126, Milan, Italy.\textsl{email:} manuela.cazzaro@unimib.it} }

%
\maketitle

\textbf{Abstract} In this work we handle with categorical (ordinal) variables and we focus on the  (in)dependence relationship  under the  marginal, conditional and context-specific perspective. If the first two are well known, the last one concerns independencies holding only in a subspace of the outcome space. We take advantage from the Hierarchical Multinomial Marginal models and provide several original results about the representation of context-specific independencies through these models. By considering the graphical aspect,  we take advantage from the chain graphical models. The resultant graphical model is a so-called "stratified" chain graphical model with labelled arcs. New Markov properties are provided. Furthermore, we consider the graphical models under the regression poit of view. Here we provide simplification of the regression parameters due to the context-specific independencies.  Finally, an application about the innovation degree of the Italian enterprises is provided. 

\section{Introduction}
\label{sec_1}

In this work we deal with categorical (ordinal) variables collected in a contingency table and we propose a model able to capture different kind of independence relationships involving ordinal variables. Different models have been proposed in the literature with the aim of describing (in)dependence relationships among the variables focusing on the independence or the dependence structure. We will refer to the \cite{bartolucci2007} Hierarchical Multinomial Marginal Models (HMMMs), that investigate the dependence structure among a set of variables. The HMMMs are specified by a set of marginals distributions together with a set of interactions defined within different marginal distributions.  Particular case of these models are the classical Log-Linear models, the \cite{bergsma2002} Marginal models, the \cite{glonek1995} Multivariate Logistic models. In particular, in this work we take advantage of the possibility of using different interactions that are significant also when we handle with ordinal variables, \cite{cazzaro2014}.

Furthermore, we will focus on the relationships among a set of categorical
(ordinal) variables under the perspective of testing, simultaneously, marginal, conditional and context-specific (CS)
independencies. The first two are well known, the (CS) independence, instead, is a conditional independence which holds only in a subspace of the outcome space. For instance, given 3 variables $X_1$, $X_2$ and $X_3$, we have $X_1\perp X_2|X_3=1$ and $X_1\not\perp X_2|X_3\neq1 $. It is interesting to study this kind of independence as it allows us to focus on the modality(ies) which discriminate and  really affect the connection among two variables.

Finally, we propose a graphical representation of all the considered independencies taking advantage of the graphical models. As a matter of fact, graphical models rotate around a system of independencies among a set of variables, and their strong benefit lies on the notable visual tool that easily represents also complex system of relationships. Different graphical models exist in literature, see \cite{lauritzen1996graphical}, \cite{whittaker2009} and \cite{wermuth2004} for an overview. Here, we start by considering a Chain Graphical (CG) model, known as type IV,  see \cite{drton2009}, adapting it according to our aims. The CG model of type IV is a naturally representation even of regression models where there are \textit{purely} response variables, \textit{purely} covariates and \textit{mixed} variables (that are covariates for some variables and responses for others).  \replaced[id=Fe]{For this reason the CG model of type IV take advantages from the so-called \textit{Multivariate Regression Markov properties}}{In fact, this kind of CG can be also see as a particular case of a Regression Chain Graphical model}, see \cite{sadeghi2016}. 

In this work, for the first time, we then integrate the CS independence in a CG model. The CS independence, under the graphical model point of view, was debated in \cite{boutilier1996}, \cite{hojsgaard2004} and \cite{nyman2016} among others. In particular, Nyman generalized the Graphical model with the so-called ``Stratified" Graphical model. Here we propose the ``Stratified" Chain Graphical (SCG) model of type IV. Furthermore, by considering the regression model represented by the CG model, the study of CS independence offers the possibility of reducing the number of parameters in complicate models. 

The work follows this structure. At first we give an overview of the HMMMs with a special attention to the representation of CS independence via HMM models, in Section \ref{sec_2}. In this section, we reach out the same results of \cite{nyman2016} by using a different approach concerning the variables coded with baseline logits. It is worthwhile to note that the known results in the literature are carried out limited to the classical log-linear models. Furthermore, in Subsection \ref{sub_loc} and \ref{sub_con}, we provide as new result, how it is possible to define CS independence by using appropriate parameters  for ordinal variables. Section \ref{sec_3} proposes the Stratified Chain Graphical (SCG) model  as a generalization of CG model of type IV. Here the Markov properties for a SCG model were provided and the admissible SCG model are discussed. In Section \ref{sec_4} we show how to parametrize a SCG model of type IV through a parametrization based on HMMM parameters. Here the original aspects are multiple. Starting from the Regression Chain Graphical model of \cite{marchetti2011} we introduce the possibility of using parameters suitable for ordinal variables, instead of the parameters based on \textit{baseline} logits. Furthermore, we provide the connection between the SGM of type IV, discussed in Section \ref{sec_3} and the HMMMs in Section \ref{sec_2}. Finally, in Section \ref{sec_5} some applications to a real dataset on the innovation status of small and medium Italian firms are shown. The conclusion is reported in Section \ref{sec_6}. All the proofs of the  theorems lie in the Appendix A in order to make more flowing the paper.

\section{Hierarchical multinomial marginal parametrization for context-specific independence}	
\label{sec_2}

Let us consider $q$ categorical variables $\mathcal{Q}=(X_1,\dots,X_q$) taking values $(i_1,$ $\dots,$ $i_q)$ in the contingency table  $\mathcal{I}=(I_1\times \dots \times I_q)$. Thus, the generic variable $X_j$ takes values in $\left\{1,\dots,I_j\right\}$. The Hierarchical multinomial Marginal Model (HMMM), introduced by \cite{bartolucci2007}, is used here in order to describe marginal, conditional and CS independence statements also when we deal with ordinal variables.

The HMMMs use parameters, henceforth HMM parameters,  that generalize the canonical log-linear parameters, by considering also the marginal distributions and possibly different coding for the logits of the variables, see \cite{cazzaro2014}. 

In particular, within a given set of variables $A$, the cells $i_A$ and $i^*_A$ represent, respectively, the $i$-th and the reference modalities of the variables in $A$ depending on the type of logits assigned to the variables on which the parameters is based.
In a \textit{baseline}, \textit{local} and \textit{continuation} logit 
 $i_{A}$ is the $i-$th modality of each variable   $\cap_{j\in A} {i_j}$. 
On the other hand, the index   $i^*_A$, in \textit{baseline} logit is $\cap_{j\in A} {I_j}$, in the \textit{local} 
logit  is    $\cap_{j\in A}\left({i_j+1}\right)$ and in the \textit{continuation} logit is $\cap_{j\in A} \sum_{s_j\geq i_j+1 }{s_j} $. Higher order parameters are obtained as contrast of logits and preserve the type of coding.  
Within a given marginal distribution $\mathcal{M}\subseteq \mathcal{Q}$, let us consider the marginal probabilities $\pi_{\mathcal{M}}$  that  is the marginal $\mathcal{M}$ probability obtained by summarizing respect to the variables $\mathcal{Q}\backslash \mathcal{M}$. Considering the HMM parameters as constrasts among the logarithms of probabilities of disjoint subsets of cells, they will  be characterized by the set $\mathcal{L}$, $\mathcal{L}\subseteq \mathcal{M}$,  of variables involved and the marginal distribution $\mathcal{M}$ where they are defined having the following form:\\
\begin{equation}
\eta_{\mathcal{L}}^{\mathcal{M}}(i_{\mathcal{L}}|i^{**}_{\mathcal{M}\backslash\mathcal{L}})=\sum_{\mathcal{J}\subseteq \mathcal{L}}(-1)^{|\mathcal{L}\backslash\mathcal{J}|}\log \pi_{\mathcal{M}}\left(i_{\mathcal{L}\backslash\mathcal{J}},i^*_{ \mathcal{J}},i^{**}_{\mathcal{M}\backslash\mathcal{L}}\right)
	\label{eq_parametri} 
\end{equation}
where $\mathcal{M}\subseteq \mathcal{Q}$ denotes the marginal table $\mathcal{I}_{\mathcal{M}}$ where the parameter is defined; $\mathcal{L}\subseteq \mathcal{M}$ is the subset of variables which the parameter refers and $i^{**}_A$ is an arbitrary cell, here the last modality $I_A$. Note that $i^{**}_{\mathcal{M}\backslash\mathcal{L}}$ select the levels of the conditioning variables. In this context, as we already cited, the reference cell involved in the  $i^{**}$ indexes will be always the last one, so following this convention we can simply denote:

\begin{equation}
\eta_{\mathcal{L}}^{\mathcal{M}}(i_{\mathcal{L}}|i^{**}_{\mathcal{M}\backslash\mathcal{L}}) =\eta_{\mathcal{L}}^{\mathcal{M}}(i_{\mathcal{L}}). \label{eq_parametri_short}
\end{equation}
Note that for each $\mathcal L$ the parameter $\eta_{\mathcal{L}}^{\mathcal{M}}(I_{\mathcal{L}})$ is trivially zero whatever the coding of the variables.  
Note that in the environment of HMMMs conditional independencies among variables can be tested by imposing to appropriate HMM parameters to be zero. For instance, given three variables $X_1$, $X_2$ and $X_3$, in order to represent the conditional independence $X_1\perp X_2|X_3$ we have that \replaced[id=Fe]{$\eta^{123}_{12}(i_{12})=\eta^{123}_{123}(i_{123})=0$ for any $i_{12}\in \mathcal{I}_{12}$ and  $i_{123}\in \mathcal{I}_{123}$, where the numbers $1$, $2$ and $3$ in the parameters refer to the variables $X_1$, $X_2$ and $X_3$, repectively.}{ $\eta^{X_1X_2X_3}_{X_1X_2}(i_{12})=\eta^{X_1X_2X_3}_{X_1X_2X_3}(i_{123})=0$ for any $i_{12}\in \mathcal{I}_{12}$ and  $i_{123}\in \mathcal{I}_{123}$}.
   \cite{bergsma2002} and \cite{bartolucci2007} proved that the above mentioned parameters provide a parameterization of the full joint probability function $\pi_{\mathcal{Q}}$ if and only if the property of hierarchy and completeness are satisfied. These two properties make sure of the smoothness of the parametrization that implies the existence of the maximum likelihood estimation. 
\begin{example}
	\label{ex_logit} Let us consider two variables $X_1$, $X_2$ collected in a $3\times3$ contingency table. In Table \ref{tab_logit} are the parameters (\ref{eq_parametri})  according to the different coding:
\scriptsize
\begin{table}[h]
	\begin{tabular}{l|ccc}
		
		type &$\eta_{1}^{12}(i_1)$   &  $\eta_{2}^{12}(i_2)$ & $\eta_{12}^{12}(i_1i_2)$   \\ 
		\hline 
		baseline & $\log\left(\frac{\pi_{33}}{\pi_{i_13}}\right)$ & $\log\left(\frac{\pi_{33}}{\pi_{3i_2}}\right)$ & $\log\left(\frac{\pi_{i_1i_2}\pi_{33}}{\pi_{i_13}\pi_{3i_2}}\right)$   \\ 
		&&&\\
		local & $\log\left(\frac{\pi_{(i_{1}+1)3}}{\pi_{i_13}}\right)$ & $\log\left(\frac{\pi_{3(i_{2}+1)}}{\pi_{3i_2}}\right)$ & $\log\left(\frac{\pi_{i_1i_2}\pi_{(i_{1}+1)(i_{2}+1)}}{\pi_{(i_{1}+1)i_2}\pi_{i_1(i_{2}+1)}}\right)$    \\ 
		&&&\\
		cont &  $\log\left(\frac{\sum_{i_1'>i_1}\pi_{(i_1')3}}{\pi_{i_13}}\right)$ & $\log\left(\frac{\sum_{i_2'>i_2}\pi_{3(i_2')}}{\pi_{3i_2}}\right)$ & $\log\left(\frac{\pi_{i_1i_2}\sum_{i_1'>i_1,i_2'>i_2}\pi_{(i_{1}')(i_{2}')}}{\sum_{i_1'>i_1}\pi_{(i_1')i_2}\sum_{i_2'>i_2}\pi_{i_1(i_2')}}\right)$   \\ 
		&&&\\		
	\end{tabular}  
	\caption{Different coding for logits and contrasts of logits.}
	\label{tab_logit}
\end{table}

\end{example}
\normalsize

The classical log-linear model is a particular case of HMMM where the parameters are all based on \textit{baseline} logit and there are only one marginal set equal to the joint distribution $\mathcal{M}=\mathcal{Q}$. \cite{nyman2016} provide the condition  to define a CS independence in classical log-linear models. Next we will reach the same condition, in a new way, for the CS independencies on the HMMMs with HMMM parameters based on baseline logit. Let us suppose we want to define a CS independence among the variables in the marginal set $\mathcal{M}$. Thus, by collecting the variables in the marginal set $\mathcal{M}$ in three subsets, supposing $A$, $B$ and $C$, we are interesting to define the following statement 
\begin{equation}
A\perp B| (C=i'_C),  \qquad i'_C\in \mathcal{K}\\
\label{nic1}
\end{equation}
where $A\cup B\cup C=\mathcal{M}$, and $i'_C$ is the vector of certain modalities of variables in $C$,  taking values in $\mathcal{K}\subset\mathcal{I}_C$, for which the conditional independence holds. 

\begin{theorem} 
	\label{T1_baseline}
	The CS independence in formula (\ref{nic1}) holds if and only if the HMM parameters, based on baseline logits, satisfy the following constraints
	\begin{equation}
	\sum_{c \in \mathcal{P}(C)
	}(-1)^{|C\backslash c|}\eta_{vc}^{\mathcal{M}}(i_v \cap i'_c)=0 \qquad 
i_v\in \mathcal{I}_v \qquad i'_c \in \left(\mathcal{K} \cap \mathcal{I}_c\right),
	\label{eq.teo1}
	\end{equation}
$\forall v\in \mathcal{V}=\left\{\left(\mathcal{P}(A)\setminus\emptyset\right)  \cup \left(  \mathcal{P}(B)\setminus \emptyset\right) \right\}$,  where $\mathcal{P}(\cdot)$ denotes the power set.
\end{theorem}

The Example \ref{ex_2.2} shows step by step how to get the constraints in formula (\ref{eq.teo1}). 
 \begin{example}
	\label{ex_2.2}Let us consider four variables collected in the marginal $\mathcal{I}_\mathcal{M}$ of dimension $3\time3\times3\times3\times3$ and let us consider the CS independence 
$X_1\perp X_2|(X_3X_4)=(1,1)$.  The HMM parameter $\eta^{1234}_{1234}(1111)$ based on \textit{baseline logit} can be decomposed as follows
\[
\begin{array}{lll}
\eta^{1234}_{1234}(1111)&=&\log\left(\frac{\pi_{3333}\pi_{1133}\pi_{1313}\pi_{3113}\pi_{1331}\pi_{3131}\pi_{3311}\pi_{1111}}{\pi_{1333}\pi_{3133}\pi_{3313}\pi_{1113}\pi_{3331}\pi_{3111}\pi_{1311}\pi_{1131}}\right)=\\
&\\
&=&\log\left(\frac{\pi_{3333}\pi_{1133}\pi_{1313}\pi_{3113}}{\pi_{1333}\pi_{3133}\pi_{3313}\pi_{1113}}\right)+ \log\left(\frac{\pi_{3333}\pi_{1133}\pi_{1331}\pi_{3131}}{\pi_{1333}\pi_{3133}\pi_{3331}\pi_{1131}}\right)+\\
&\\
&&- \log\left(\frac{\pi_{3333}\pi_{1133}}{\pi_{1333}\pi_{3133}}\right)+
\log\left(\frac{\pi_{3311}\pi_{1111}}{\pi_{3111}\pi_{1311}}\right)=\\
&\\
&=&(-1)^{|\{4\}|+1}\eta^{1234}_{123}(111)+(-1)^{|\{3\}|+1}\eta^{1234}_{124}(111)+\\
&&\\
&&(-1)^{|\{3,4\}|+1}\eta^{1234}_{12}(11)+(-1)^{|\{3,4\}|}\eta^{1234}_{12}(11|11).
\end{array}
\]
From the CS independence we have that $\eta^{1234}_{12}(11|11)=0$ and by shifting the right hand side on the left we get:
\[
\eta^{1234}_{1234}(1111)-\eta^{1234}_{123}(111)-\eta^{1234}_{124}(111)+\eta^{1234}_{12}(11)=0
\]
The same equivalence holds for $\eta^{1234}_{1234}(1211)$, $\eta^{1234}_{1234}(2111)$ and $\eta^{1234}_{1234}(2211)$.

Note that, having the CS independence $ X_1\perp X_2|(X_3X_4) =(1,3)$, the constraints involving the variables $X_4$ at the fourth modality are zero by definition, thus formula (\ref{eq.teo1}) becomes

\[
\begin{array}{lll}
-\eta^{1234}_{123}(111)+\eta^{1234}_{12}(11)=0\\
\\
-\eta^{1234}_{123}(121)+\eta^{1234}_{12}(12)=0\\
\\
-\eta^{1234}_{123}(211)+\eta^{1234}_{12}(21)=0\\
\\
-\eta^{1234}_{123}(221)+\eta^{1234}_{12}(22)=0\\
\end{array}
\]
\added[id=Fe]{INVECE DELLA FORMULA PRECEDENTE MEGLIO QUESTA?}
\[-\eta^{1234}_{123}(i_{123})+\eta^{1234}_{12}(i_{12})=0\]
\added[id=Fe]{where $i_{123}\in\left\{(111),(121),(211),(221)\right\}$ and $i_{12}\in\left\{(11),(12),(21),(22)\right\}$.}
\end{example}
\begin{remark} 
	\label{rem_1}If in the CS statement in formula (\ref{nic1})   $\mathcal{K}=\mathcal{I}_C$, then the constraints in formula  (\ref{eq.teo1}) satisfy the conditional independence $A\perp B|C$.\\
\end{remark}

From Remark \ref{rem_1} comes that the CS independence $A\perp B|D(C=i'_C)$, for $i'_C\in \mathcal{K}$, matches with the CS independence \replaced[id=Fe]{$A\perp B|(DC=i'_{DC})$, where $i'_{DC}=i'_D\cap i'_C$,}{ $A\perp B|(DC=i_Di_C)$}, for $i'_C\in \mathcal{K}$ and $i'_D\in \mathcal{I}_D$. Henceforth, this situations will be described as $A\perp B|(DC)=(*,i'_C)$ where the asterisk denotes we refer to all modalities.
\begin{remark} \label{rim_df} Given a CS statement as in formula (\ref{nic1}), the number of constraints imposed at a saturated log-linear model are $\left[\left(\prod_{j\in(A\cup B)}I_j\right)-1\right] \times |\mathcal{K}|$.
\end{remark}

As mentioned before, the aim of this work is to provide a model able to represent the CS independence statements by considering also the ordinal variables. When we handle with ordinal variables, \textit{baseline} logits are no longer appropriate. The \textit{local}, \textit{continuation} or \textit{reverse} approaches are more suitable. The following subsections deal with these logits. 

\subsection{Constraints on HMM parameters based on \textit{local} logit}
\label{sub_loc}
Let us suppose that the conditional set in (\ref{nic1}) is composed only of ordinal variables and we use parameters based on \textit{local} logits to code these ones, then the CS independence can be described by Theorem \ref{T2_local}.

\begin{theorem}
	\label{T2_local}
		The CS independence in formula (\ref{nic1}) holds if and only if the HMM parameters based on \textit{local} logits satisfy the following constraints
	\begin{equation}
	\sum_{c \in \mathcal{P}(C)
}(-1)^{|C\backslash c|}	\sum_{i_c \geq i'_c
				}\eta_{vc}^{\mathcal{M}}(i_{vc})=0 
	\label{nic2}
	\end{equation}
	$\forall v\in \mathcal{V}$ where $\mathcal{V}= \left\{\left(\mathcal{P}(A)\setminus\emptyset\right)  \cup \left(  \mathcal{P}(B)\setminus \emptyset\right) \right\}$, 	$i_{vc}=i_v \cap i_c$, $\forall i_v\in \mathcal{I}_v$ and $\forall i'_c \in \left(\mathcal{K} \cap \mathcal{I}_c\right)$.   
\end{theorem}

\begin{example}
	\label{ex_local1}
	Let us consider the  case of three variables collected in a $2\times2\times4$ contingency table. If we want to consider the CS independence $X_1\perp X_2|X_3=2$ where all the variables are coded with \textit{local} approach in the parameters we consider the decomposition in formula (\ref{nic2}):
	\[
	\begin{array}{ll}
	e^{\left(\left(\eta^{123}_{123}(112)+\eta^{123}_{123}(113)\right)-\eta^{123}_{12}(11)\right)}&=\left(\frac{\pi_{223}\pi_{113}\pi_{122}\pi_{212}}{\pi_{123}\pi_{213}\pi_{222}\pi_{112}}\right)\left(\frac{\pi_{224}\pi_{114}\pi_{123}\pi_{213}}{\pi_{124}\pi_{214}\pi_{223}\pi_{113}}\right)\left(\frac{\pi_{124}\pi_{214}}{\pi_{224}\pi_{114}}\right)=\\	
	&\\
	&=\frac{\pi_{122}\pi_{212}}{\pi_{222}\pi_{112}}
	\end{array}
	\]\\
\replaced[id=Fe]{	that becomes equal to $1$ when the CS independence holds, thus the $\log$ of the previuos fraction is equal to $0$.}{	that when the CS independence holds become equal to $1$, thus the $\log$ is equal to $0$.}
\end{example}

Until now we consider the CS independence like in formula (\ref{nic1}), but when we handle with ordinal variables a more interesting specification of CS independence is 
\begin{equation}
A\perp B| C\geq i'_C, \qquad  i'_C\in \mathcal{K}
\label{nic_cs_1}
\end{equation}
or
\begin{equation}
A\perp B| C\leq i'_C, \qquad i'_C \in \mathcal{K}
\label{nic_cs_2}
\end{equation}
where in this case the class $\mathcal{K}$ is composed of only one cell $i'_C$ and the CS independence must hold for all modalities of variables in $C$ greater(lower) than or equal to the cell in $\mathcal{K}$. 
Obviously, if the constraints in Theorem \ref{T2_local} are satisfied for each $i_C\geq i'_C$ ($i'_C\leq i_C$), then the (\ref{nic_cs_1}) (or (\ref{nic_cs_2})) holds too. But in the case of \textit{ local} parameters, there is a easiest way to define the CS independence in formula (\ref{nic_cs_1}), as shown in Corollary \ref{cor1}.
\begin{corol}
	\label{cor1}
The CS independence in formula (\ref{nic_cs_1}) holds if and only if the HMM parameters based on \textit{local} logits satisfy the following constraints:
\begin{equation}
\eta^{\mathcal{M}}_{vc}(i_{vc})=0 \qquad 	i_{vc}=i_v \cap i_c \qquad  i_c \geq i'_c  \qquad i'_c \in \left(\mathcal{K} \cap \mathcal{I}_c\right) \qquad  i_v\in \mathcal{I}_v
\label{eq_cor1}
\end{equation}
$\forall v\in \mathcal{V}$ \added[id=Fe]{where $\mathcal{V}= \left\{\left(\mathcal{P}(A)\setminus\emptyset\right)  \cup \left(  \mathcal{P}(B)\setminus \emptyset\right) \right\}$,} and $\forall c\in \mathcal{P}(C)$ with $c \neq \emptyset$. 
\end{corol}
\begin{example}
\label{ex_local_2} From Example \ref{ex_local1} let us consider a marginal set $\mathcal{M}=(X_1, X_2, X_3)$. The CS independence $X_1\perp X_2 |X_3\geq 2$ holds if
\[
\begin{array}{lll}
\eta^{123}_{12}(11)=0 &\quad \eta^{123}_{123}(112)=0&\quad \eta^{123}_{123}(113)=0.\\
\end{array}\] 
\end{example}

\subsection{Constraints on parameters based on \textit{continuation} logit}
\label{sub_con}

As it is shown in Table \ref{tab_logit}, the parameters based on \textit{continuation} logits 
involve also sum of probabilities. This make impossible to explicit constraints to define the CS independence as defined in formula (\ref{nic1}). However, since this kind of parametrization is adopted when the variables are ordinal, it is helpful also to consider the particular cases displayed in formula (\ref{nic_cs_1}) and (\ref{nic_cs_2}). In this section we deal with these questions.\\

\begin{theorem}
	\label{teo3}
	The CS independence in formula (\ref{nic_cs_1}) holds if and only if the HMM parameters based on \textit{continuation} logits satisfy the following constraints:
\begin{equation}
\eta^{\mathcal{M}}_{vc}(i_{vc})=0 \qquad 	i_{vc}=i_v \cap i_c \qquad  i_c \geq i'_c  \qquad i'_c \in \left(\mathcal{K} \cap \mathcal{I}_c\right) \qquad  i_v\in \mathcal{I}_v
\label{eq_cor1bis}
\end{equation}
$\forall v\in \mathcal{V}$, \added[id=Fe]{where $\mathcal{V}= \left\{\left(\mathcal{P}(A)\setminus\emptyset\right)  \cup \left(  \mathcal{P}(B)\setminus \emptyset\right) \right\}$,} and $\forall c\in \mathcal{P}(C)$ with $c \neq \emptyset$. 
\end{theorem}

\begin{example}
	\label{ex_continuation}
	Let us consider the situation described in the Example \ref{ex_local_2} but with parameters based on \textit{continuation} logits.
	The parameters involved in Theorem \ref{teo3} are $\eta_{12}^{123}(11)$,  $\eta_{123}^{123}(112)$ and $\eta_{123}^{123}(113)$. In particular, the first is
	\[
\eta_{12}^{123}(11)=\log\left(\frac{\pi_{114}\pi_{224}}{\pi_{124}\pi_{214}}\right).
	\]
	Note that, $X_1\perp X_2|X_3\geq 2$ implies  $X_1\perp X_2|X_3=4$. Then the previous parameter is equal to zero.

About the second parameter, we have:
		\[
\eta_{123}^{123}(112)=\log\left(\frac{\left(\pi_{223}+\pi_{224}\right)\left(\pi_{113}+\pi_{114}\right)\left(\pi_{122}\right)\left(\pi_{212}\right)}{\left(\pi_{123}+\pi_{124}\right)\left(\pi_{213}+\pi_{214}\right)\left(\pi_{222}\right)\left(\pi_{112}\right)}
\right).
	\]
Since the variable $X_3$ appears only with modalities $2, 3 $ and $4$ for which the CS independence holds, then we get 
	that even this parameter is null. In the same way we progress for the third parameter that is equal to zero.
\end{example}

\begin{remark}
When we are interested in defining a CS independence as expressed in formula (\ref{nic_cs_2}), we can proceed in an analogous way previously  sorting in a descending order the modalities of the interest variable. This corresponds to the \textit{reverse continuation} coding of the variable.
\end{remark}	
	 Thus, if, for instance, we are interested in checking if a CS independence between two variables holds when the population is young or adult against old, we can sort the modalities of the variable \textit{Age} in the reverse order $\left\{Old, Adult, Young\right\}$ and then consider the CS independence in formula (\ref{nic_cs_1}).\\	 
In general, we can decide to codify the variables heterogeneously, with different kinds of logits, in order to suit the nature of the variables. However, as it is shown in this section, the constraints required to define CS independence statements depend on the type of logits used to code the variables in the conditional set. Here we present an example in order to show how to apply the different theorems when we handle with variables coded with different type of logits.

\begin{example}
\label{ex_misto}
Let us consider a marginal set $\mathcal{M}$ composed of 4 variables collected in a $2\times 2\times 4\times4$ contingency table $\mathcal{I}_\mathcal{M}$. 
We codify the variables with \textit{baseline}, \textit{baseline}, \textit{local} and \textit{continuation} logits, respectively. We are interested in checking the CS independence $X_1\perp X_2| X_3X_4\geq (2,2)$  that means that the CS independence must hold when the variables $X_3$ and $X_4$ assume, respectively, the values $X_3\geq 2$ and $X_4\geq 2$ that is the levels $\left\{(2,2);(2,3);(3,2);(3,3)\right\}$. In this case, noting that the variables in the conditioning set are coded with the local and the continuation logits, the results due to Corollary \ref{cor1} and Theorem \ref{teo3} imply that the following parameters, involving the conditioning variables with values greater or equal to $(2,2)$, have to be zero, how effectively is:
\[
\begin{array}{ll}
\eta_{1234}(1122)=&\log\left(\frac{\left(\pi_{1122}\right)\left(\pi_{2222}\right)\left(\pi_{2132}\right)\left(\pi_{2123}+\pi_{2124}\right)\left(\pi_{1232}\right)\left(\pi_{1223}+\pi_{1224}\right)\left(\pi_{1133}+\pi_{1134}\right)\left(\pi_{2233}+\pi_{2234}\right)}{\left(\pi_{2122}\right)\left(\pi_{1222}\right)\left(\pi_{1132}\right)\left(\pi_{1123}+\pi_{1124}\right)\left(\pi_{2232}\right)\left(\pi_{2223}+\pi_{2224}\right)\left(\pi_{2133}+\pi_{2134}\right)\left(\pi_{1233}+\pi_{1234}\right)}\right)=0\\
&\\
\eta_{1234}(1132)=&\log\left(\frac{\left(\pi_{1132}\right)\left(\pi_{2232}\right)\left(\pi_{2142}\right)\left(\pi_{2133}+\pi_{2134}\right)\left(\pi_{1242}\right)\left(\pi_{1233}+\pi_{1234}\right)\left(\pi_{1143}+\pi_{1144}\right)\left(\pi_{2243}+\pi_{2244}\right)}{\left(\pi_{2132}\right)\left(\pi_{1232}\right)\left(\pi_{1142}\right)\left(\pi_{1133}+\pi_{1134}\right)\left(\pi_{2242}\right)\left(\pi_{2233}+\pi_{2234}\right)\left(\pi_{2143}+\pi_{2144}\right)\left(\pi_{1243}+\pi_{1244}\right)}\right)=0\\
&\\
\eta_{1234}(1123)=&\log\left(\frac{\left(\pi_{1123}\right)\left(\pi_{2223}\right)\left(\pi_{2133}\right)\left(\pi_{2124}\right)\left(\pi_{1233}\right)\left(\pi_{1224}\right)\left(\pi_{1134}\right)\left(\pi_{2234}\right)}{\left(\pi_{2123}\right)\left(\pi_{1223}\right)\left(\pi_{1133}\right)\left(\pi_{1124}\right)\left(\pi_{2233}\right)\left(\pi_{2224}\right)\left(\pi_{2134}\right)\left(\pi_{1234}\right)}\right)=0\\
&\\
\eta_{1234}(1133)=&\log\left(\frac{\left(\pi_{1133}\right)\left(\pi_{2233}\right)\left(\pi_{2143}\right)\left(\pi_{2134}\right)\left(\pi_{1243}\right)\left(\pi_{1234}\right)\left(\pi_{1144}\right)\left(\pi_{2244}\right)}{\left(\pi_{2133}\right)\left(\pi_{1233}\right)\left(\pi_{1143}\right)\left(\pi_{1134}\right)\left(\pi_{2243}\right)\left(\pi_{2234}\right)\left(\pi_{2144}\right)\left(\pi_{1244}\right)}\right)=0\\
&\\
\eta_{123}(113)=&\log\left(\frac{\left(\pi_{2134}\right)\left(\pi_{1234}\right)\left(\pi_{1144}\right)\left(\pi_{2244}\right)}{\left(\pi_{1134}\right)\left(\pi_{2234}\right)\left(\pi_{2144}\right)\left(\pi_{1244}\right)}\right)=0\\
&\\
\eta_{124}(112)=&\log\left(\frac{\left(\pi_{2142}\right)\left(\pi_{1242}\right)\left(\pi_{1143}+\pi_{1144}\right)\left(\pi_{2243}+\pi_{2244}\right)}{\left(\pi_{1142}\right)\left(\pi_{2242}\right)\left(\pi_{2143}+\pi_{2144}\right)\left(\pi_{1243}+\pi_{1244}\right)}\right)=0\\
&\\
\eta_{124}(113)=&\log\left(\frac{\left(\pi_{2143}\right)\left(\pi_{1243}\right)\left(\pi_{1144}\right)\left(\pi_{2244}\right)}{\left(\pi_{1143}\right)\left(\pi_{2243}\right)\left(\pi_{2144}\right)\left(\pi_{1244}\right)}\right)=0\\
&\\
\eta_{12}(11)=&\log\left(\frac{\left(\pi_{1144}\right)\left(\pi_{2244}\right)}{\left(\pi_{2144}\right)\left(\pi_{1244}\right)}\right)=0

\end{array}\]
The same holds for the remaining modalities of $X_1X_2$.
\end{example}

\section{Stratified Chain Graphical models of type IV}
\label{sec_3}
In this section we will handle with the Chain Graphical models, thus a brief review on these tools is necessary.\\
 Formally, a \textit{Chain Graph} (CG) is a graph $G=\left\{V,E\right\}$ that is a collection of vertices and edges, with  both directed and undirected arcs in  $E$ and without any directed or semi-directed cycle. Two vertices  linked by an undirected arc are \textit{adjacent}.  Given a set $A$ of vertices, the \textit{neighbour} of $A$, $nb(A)$, is the set of vertices adjacent to at least one vertex in $A$. The \textit{neighbourhood}, $Nb(A)$,  add to the neighbour set the $A$ itself: $nb(A)\cup A$. A set $A$ is called \textit{non connected} if there is not a path that links all the vertices in the set. The set of vertices from which directed arcs start, pointing all to $A$, is called \textit{parent} set, $pa_G(A)$.  
 A CG is characterized by the so-called \textit{chain components}, denoted by $T_1,....,T_s$, where the vertices are partitioned according to the following conventions. Vertices linked by undirected arcs must belong to the same component and vertices linked by directed arcs must belong to different components. The set of components from which start at least one directed arc pointing to the component $T_h$ is called parent component, $pa_D(T_h)$. Finally the \textit{non descendant} of the component $T_h$, $nd(T_h)$, is composed of the components that cannot be reached from $T_h$ by a direct path.\\
The Chain Graphical Model (CGM) is a model of conditional and marginal independencies represented by a CG where the variables are represented by vertices and the relationships between variables through arcs. This kind of model is useful when  the analysed variables follow an inherent explicative order such as some variables are  explicative of other variables which can be in turn explicative variables for other ones. Thus, the partition of the vertices in components comes naturally according to the variables which vertices represent.\\ 
As shown by \cite{drton2009}, there are different rules to extract a list of independencies between variables from a CG. These rules are called Markov Properties and characterize 4 types of CGM. In this work we take advantage from the CGM of type IV \added[id=Fe]{also known as multivariate regression Markov Properties}, \cite{sadeghi2014}. 

\begin{defi}
	Given a CG, the Markov Properties of type IV to extract a list of conditional and marginal independencies are:
\begin{equation}
\begin{array}{lll}
\textsf{C1)}\quad &T_h\perp nd(T_h)|pa_D(T_h),\quad  &h=1,\dots,s;\\
\textsf{C2)}\quad & A\perp T_h\backslash Nb(A)|pa_D(T_h), \quad & h=1,\dots,s, \qquad A\subset T_h;\\
\textsf{C3)}\quad &A\perp pa_D(T_h)\backslash pa_G(A)|pa_G(A), \quad & h=1,\dots,s, \qquad A\subset T_h.\\
\end{array}
\label{MP_IV}
\end{equation}
\end{defi}
Note that this type of CGM identifies the independencies between variables involved in the same component as  marginal independencies. \\

\begin{figure}[!ht]
	\begin{center}
		
		\begin{minipage}[t]{0.45\textwidth}
			\begin{center}
			\includegraphics[width=4cm]{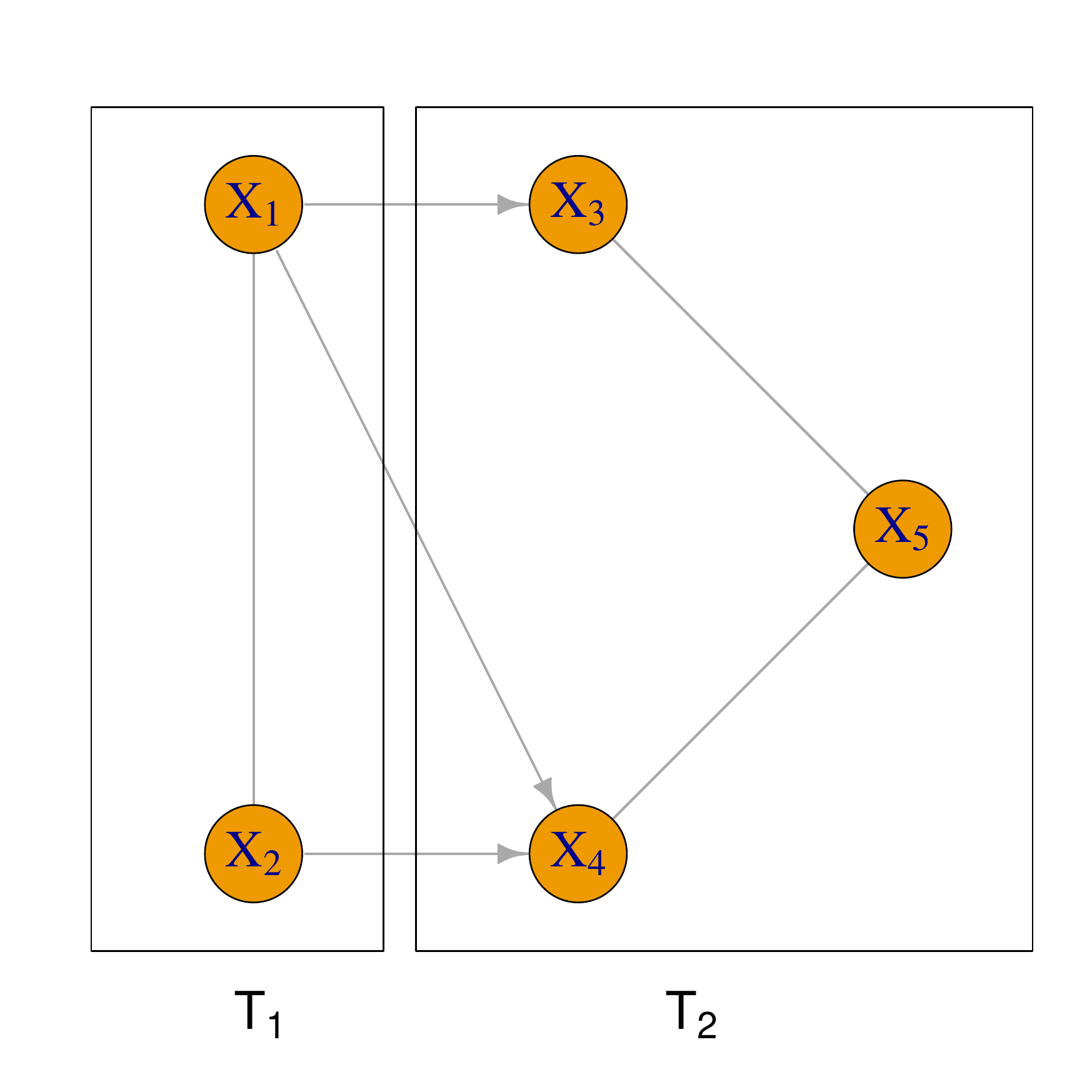}\\
			 (a)
		 	\end{center}
				\end{minipage}
		\begin{minipage}[t]{0.45\textwidth}
						\begin{center}
			\includegraphics[width=4cm]{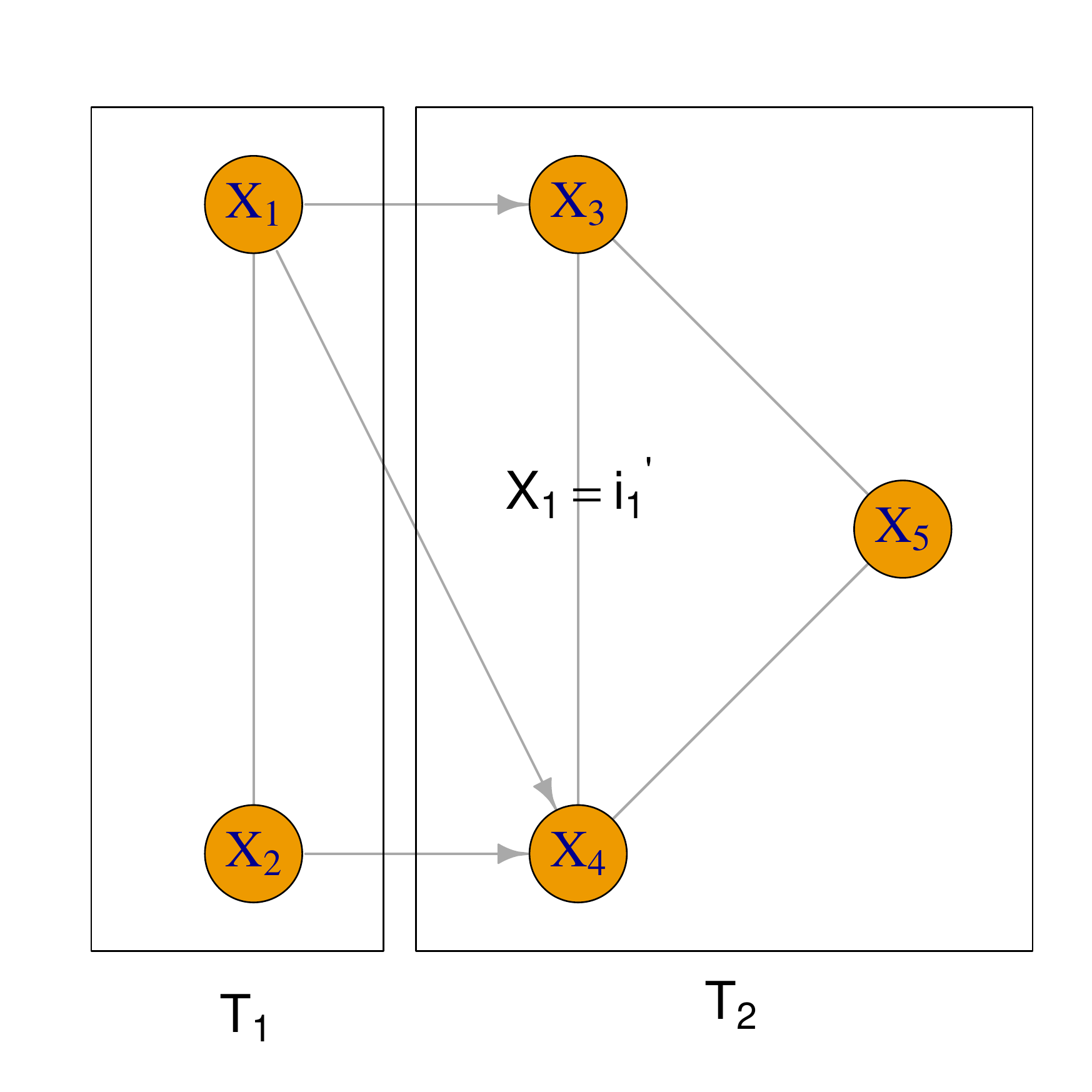}\\
					(b)
					\end{center}
			 \end{minipage}
		\caption{\label{nicolussi:fig1} \replaced[id=Fe]{CG (on the left) and  SCG  with the stratum $\mathcal{K}^{12}_{34}=\left\{(i_1',*)\right\}$ (on the right), both with components $T_1=\left(1,2\right)$ and $T_2=\left(3,4,5 \right)$.}{CG (on the left) and  SCG  with the stratum $S_{34}=\left\{12=(i_1',*)\right\}$ (on the right), both with components $T_1=\left(1,2\right)$ and $T_2=\left(3,4,5 \right)$.}}
	\end{center}
\end{figure}

\begin{example}
Let us consider the CG in Figure \ref{nicolussi:fig1} where we can recognize two components: $T_1=\left(1, 2\right)$ and $T_2=\left(3,4,5 \right)$. By applying the Markov Properties in (\ref{MP_IV}), focusing on Figure \ref{nicolussi:fig1} (a), we get the following list of independencies: $3\perp 4|12$, $3\perp 2|1$ and $5\perp 12$. 
\label{ex_CG4}
\end{example}

 In order to take into account the CS independencies, we propose the Stratified Chain Graphical Models (SCGMs) as an extension of the Stratified Graphical Models (SGMs) proposed by \cite{nyman2016}. Similarly to SGM, we denote the CS independencies throw labeled arcs, denoted as \textit{stratum}, $S$. \added[id=Fe]{The example \ref{ex_stratum} shows briefely how to interpret the \textit{stratum} in the SCGM before the tecnical explanation below.}
 
 \begin{example}
 	\label{ex_stratum} 
 	\added[id=Fe]{In Figure \ref{nicolussi:fig1} (b), we have the labelled arc between the nodes $3$ and $4$ which reports the modality $i_1'$ of the variable $1$. This \textit{stratum}  stands for $3 \perp 4|12=(i_1',*)$, where the asterisk denotes that the independence holds for any modality of $2$.} 
 \end{example}

\begin{figure}[h]
	\centering
\includegraphics[width=4cm]{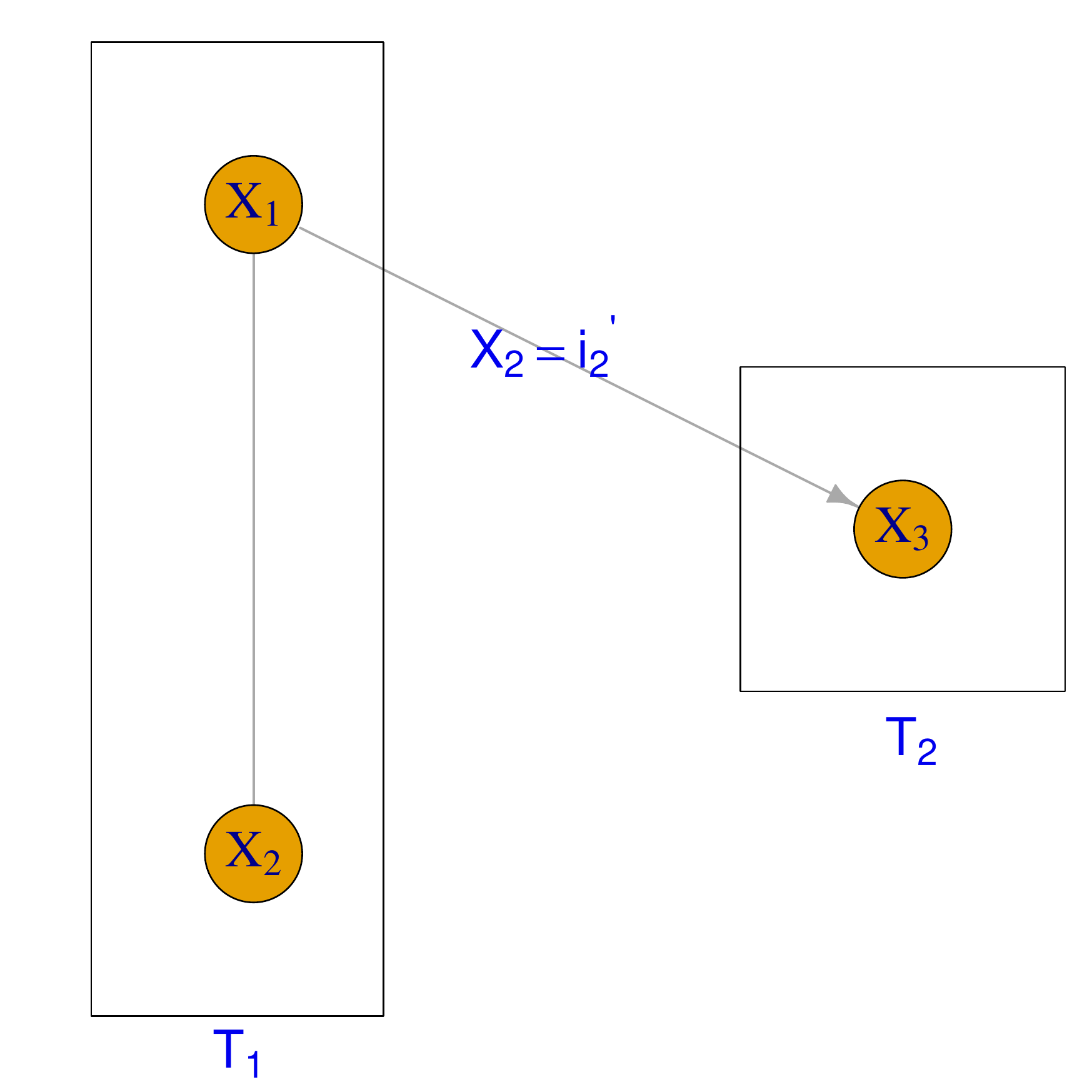}
\caption{SCG with components $T_1=\left(1,2\right)$ and $T_2=\left(3\right)$ with a non representable \textit{stratum}.}
\label{no_admit}
\end{figure}

Formally, a SCG is defined by three sets, the \added[id=Fe]{one of} vertices $V$, the \added[id=Fe]{one of} edges $E$ and the \added[id=Fe]{one of} stratum $S$ which denotes the labelled arcs. \replaced[id=Fe]{In particular each element of S is a stratum $\mathcal{K}^C_{\gamma,\delta}=\left\{i'_C:\gamma\perp \delta |C=i'_C\right\}$.
	 which refers to a pair of vertices $(\gamma,\delta)$ and which reports the list of modalities of the veriables in $C$  according to which the arc is missed (and the CS independence holds).}{In particular each stratum $S_{\gamma,\delta}={S=i_{S}}$ refers to a pair of vertices $(\gamma,\delta)$ and reports the list of modalities ($i_S$) of the variables in $S$ according to which the arc is missed} \added[id=Fe]{ Since we are using the SCGM as extension of CGM of type IV, the set $C$, is always included  or equal to the set of parents  of vertices $(\gamma, \delta)$}. \deleted[id=Fe]{Note that in Figure \ref{nicolussi:fig1} (b) if the arc between $3$ and $4$ was missed, we have the independence $3\perp 4|12$, however on the labelled arc are reported only the modalities of the variable $1$, this means that any modality of $2$ produces an independence, thus is unnecessary to write it.} 
\begin{defi} Given a SCG, the \added[id=Fe]{Stratified} Markov Properties to extract a list of conditional, marginal and CS independencies are
	\begin{equation}
	\begin{array}{llll}
	\textsf{C1)}\quad &T_h\perp nd(T_h)|pa_D(T_h),\quad  &h=1,\dots,s;&\\
	\textsf{CS2)}\quad &\gamma\perp \delta |pa_D(T_h)=i'_{pa_D(T_h)};\quad &i'_{pa_D(T_h)}\in\mathcal{K}_{\gamma,\delta}^{pa_D(T_h)} \qquad &\text{if}\quad \gamma,\delta \in T_h;\\ 
	\textsf{CS3)}\quad &\gamma\perp \delta |S_{\gamma,\delta}=pa_G(\gamma)=i'_{pa_G(\gamma)} \quad &i'_{pa_G(\gamma)}\in\mathcal{K}_{\gamma,\delta}^{pa_G(\gamma)} \qquad &\text{if}\quad \gamma\in T_h, \delta\in pa_D(T_h)\backslash pa_G(\gamma) \\
	\end{array}
	\label{SMP_IV}
	\end{equation}
	

where the \textbf{C1)} is equal to the rule \textbf{C1)} in formula (\ref{MP_IV}) and \textbf{CS2)} and \textbf{CS3)} are a generalization of the remaining rules in formula  (\ref{MP_IV}).
\end{defi}
\deleted[id=Fe]{In the conditional set of both \textbf{CS2)} and \textbf{CS3)}, we have that $S_{\gamma,\delta}\subseteq pa_D(T_h)$ and $S_{\gamma,\delta}\subseteq pa_G(\gamma)$. }
\added[id=Fe]{
When in the conditional set of both \textbf{CS2)} and \textbf{CS3)}, the \textit{stratum} $\mathcal{K}_{\gamma,\delta}^{C}$ coincide with  $\mathcal{I}_C$, we are handling with a conditional indpendence and the \textit{stratum} is unnecessary. In this case we bring back to the ``pairwise" Markov properties} for a CGM of type IV  \added[id=Fe]{as listed in formula (2) of} \cite{marchetti2011} \added[id=Fe]{that are equivalent to the ones in formula (\ref{MP_IV}).}
\deleted[id=Fe]{On the other hand, when any variable $c$ in $C$  assumes a subset of modalities of $\mathcal{I}_c$,  the asterisk in the CS independencies come from the \textbf{CS2)} and \textbf{CS3)}, drops out.}\\
Graphically, a \textit{stratum} can be an undirected labelled arc \added[id=Fe]{the case of \textbf{(CS2)}}, or a directed labelled arc, \added[id=Fe]{the case pf \textbf{CS3}}. \deleted[id=Fe]{, but the variables in the \textit{stratum} belong only to the parent set.} However, not any possible \textit{stratum} is admitted in the SCG model. Let us consider, for instance, the graph in Figure \ref{no_admit}, we have the conditional independence $3\perp 2| 1$ but, at the same time we have the CS independence $3\perp 1|2=i_2'$. In the conditional independence we declare that the variable $2$ does not affect variable $3$ for any modality of $1$ but, in the CS independence we affirm that the variable  $2$ discriminates the relationship between $1$ and $3$, thus it has some effect on the variable $3$.
\cite{nyman2016} dealt with this situation and, in their Theorem 2,  they give  the condition for the existence of a stratum that is summarized in the following remark.

\begin{remark}\label{rem_stratum}
 Given a SCG, with at least one stratum $\mathcal{K}_{\gamma,\delta}^{C}\not\equiv \mathcal{I}_{C}$, then the variables in $S$ must be adjacent or parents of both $\gamma$ and $\delta$.
\end{remark}

\begin{example}
	\label{ex_stratum2} 
	\added[id=Fe]{(\textit{continuation of example \ref{ex_stratum}})}
		In Figure \ref{nicolussi:fig1} (b), we have the stratum $\mathcal{K}^{12}_{34}=\left\{(i_1',*)\right\}$. Thus, by applying the \textbf{CS2)}, the statement $3\perp 4|12=(i_1',*)$ comes. 
	Note that,  the variable $1$ involved by the stratum belongs to both $pa_G(3)$ and $pa_G(4)$ and the previous remark \ref{rem_stratum} holds.
\end{example}
 
\section{Regression model with context specific independencies}
\label{sec_4}
In Section \ref{sec_2} the main results about the HMM parametrization are discussed, while in Section \ref{sec_3} a graphical model for different kind of independencies is presented. In this section we connect these two models and we show how we can parametrize a SCGM through a HMMM.\\ 
As mentioned in Section \ref{sec_3}, the approach of CGMs seems natural  to explain  the effect of some variables (covariates) on a set of dependence variables that can be in turn covariates for other dependence variables. Thus, it is appropriate to collect the variables in the components according to this purpose and, by focusing on each component $T_h$, we consider as covariates of $T_h$ the variables in $pa_D(T_h)$.  
The CGM of type IV admits to simplify the regression statements by using a marginal approach for the variables in the same components, as it is shown by \cite{marchetti2011}.  Here we want to improve this \replaced[id=Fe]{multivariate regression model based on SCGMs}{Chain Regression model} by considering ordinal variables coded by \textit{local} logits and then by simplifying the regression equations thanks to the CS independencies.
As it is shown in  \cite{marchetti2008}, \cite{rudas2010} and \cite{Nicolussi2013} the CGM of type IV can be parametrized by using the HMMMs with the appropriate hierarchical marginal sets $\mathcal{H}=\left\{\mathcal{H}_1,\mathcal{H}_2\right\}$ where
\begin{equation}
\begin{array}{ll}
\mathcal{H}_1=&\left\{(pa_D(T_h)\cup A),\,h=1,\dots,s;\,A\subseteq T_h\right\}\\
\mathcal{H}_2=&\left\{(nd(T_h)\cup T_h),\,h=1,\dots,s \right\}.
\end{array}
\label{marginal}
\end{equation}
These two classes must be put together in $\mathcal{H}$ so that if $j< i$ then $\mathcal{M}_i \not\subseteq \mathcal{M}_j$.
Then, focusing on each group of dependent variables, we define the HMM parameters (\ref{eq_parametri_short})  evaluated in each conditional distribution of the covariates. That means, for each set of dependent variables $A\subseteq T_h$, we define the parameters $\eta^{A\cup pa_D(T_h)}_{A}(i_A|i_{pa_D(T_h)})$ evaluated in any values $i_{pa_D(T_h)}\in\mathcal{I}_{pa_D(T_h)}$ of the covariates $pa_D(T_h)$. All these parameters can be expressed as combination of regression parameters as follows.  

\begin{defi}
	\label{def_3}
Given a SCGM,  \textit{regression parameters} are given by
 \begin{equation}
\eta^{A\cup pa_D(T_h)}_{A}(i_A|i_{pa_D(T_h)})= \sum_{t\subseteq pa_D(T_h)} \beta_{t}^{A}(i_t) \qquad  \forall h=1,\dots,s\qquad  A\subseteq T_{h}.
	\label{eq:regression_param}
\end{equation}
\end{defi}
\begin{theorem}
\label{teo:param}
The parameters $\beta_{t}^{A}(i_t)$  in the regression model (\ref{eq:regression_param}), are the HMM parameters based on \textit{baseline} or \textit{local} logit
\begin{equation}
\label{formula}
 \beta^A_{t}(i_t)=(-1)^{|pa_D(T_h)\backslash t|}\eta^{(A\cup pa_D(T_h))}_{tA}(i_{tA}|i^{*}_{pa_D(T_h)\backslash t})
\end{equation}
$\forall t\subseteq pa_D(T_h)\neq \emptyset$ \added[id=Fe]{and where $i_{tA}=i_t\cap i_A$}.
\end{theorem}

\begin{example}
	Let us consider the CGM in Figure \ref{nicolussi:fig1} (a) where there are two components. The first is composed of the purely dependent variables $3$, $5$ and $6$ while in the second there are the covariates $1$ and $2$. Thus, according to the formula  (\ref{marginal}), the marginal sets take values in $\left\{(123),(124),\right.$ $\left.(125),(1234),(1235),(1245),(12345)\right\}$. By focusing on the dependent variable $4$, we can express the regression model as follows 

	\[
\eta^{123}_{4}(i_4|i_{12})=\beta_{\emptyset}^{4}+\beta_{1}(i_1)^{4}+\beta_{2}^{4}(i_2)+\beta_{12}^{4}(i_{12})
\]
 $\forall  i_{12}\in \mathcal{I}_{12} $ and $i_4\in\left\{1,\dots,I_4-1\right\} $ because  when $i_4= I_4$  the parameter is zero by definition. By applying Corollary \ref{cor1_APP} in Appendix \ref{appendix_1}, 
 we see that the $\beta$ parameters are 
 \[
 \begin{array}{lll}
 \beta^4_{\emptyset}&=&\eta^{(124)}_{4}(i_{4}|i^{*}_{12})\\ \beta^4_{1}(i_1)&=-&\eta^{(124)}_{14}(i_{14}|i^{*}_{2})\\
  \beta^D_{2}(i_2)&=-&\eta^{(124)}_{24}(i_{24}|i^{*}_{1})\\
   \beta^D_{12}(i_{12})&=&\eta^{(124)}_{124}(i_{124}). 
 \end{array}
 \]
\end{example}
The parameters in formula (\ref{eq:regression_param}) are able to explain the relationship between variables in $T_h\cup pa_D(T_h)$ for each $h=1,\dots,s$. The remaining relationships between variables belonging to disjointed components can be described by the HMM parameters
\begin{equation}
\eta^{T_h\cup nd(T_h)}_{AB}(i_{AB})
\label{mix}
\end{equation} 
\added[id=Fe]{where $h=1,\dots,s$, $A\subseteq T_h$ and $B\subseteq nd(T_h)$ such that $B\cap (nd(T_h)\backslash pa_D(T_h)\neq \emptyset$}.
\begin{theorem}The regression parameters in formula (\ref{eq:regression_param}) and the HMM parameters in formula (\ref{mix}) are a 1:1 function (a reparametrization) of the HMM parameters $\eta_{\mathcal{L}}^{\mathcal{M}}$, $\forall \mathcal{L}\in\mathcal{P}(\mathcal{Q})$ and $\forall \mathcal{M}\in\mathcal{H}$.	\label{parametrizz}
\end{theorem}
Now let us to consider the SCGM as presented in Section \ref{sec_3}. The previous considerations about the parametrization still holds and the following theorem explains how to constrain the HMM parameters according to the SGCM. 
\begin{theorem}
\label{regression_constraints}
A SCGM that obeys to the \added[id=Fe]{Stratified} Markov Properties of type IV in (\ref{SMP_IV}) can be parametrized as follows:
	\begin{itemize}
	\item[i)]Each (C1) holds iif the   HMM parameters 
in formula (\ref{mix}) are equal to zero;\\
		\item[ii)] Each (CS2)  holds iif the   regression parameters 
		in formula (\ref{eq:regression_param}) satisfy  
$\eta^{\mathcal{L}\cup pa_D(T_h)}_{\mathcal{L}}(i_{\mathcal{L}}|i'_{pa_D(T_h)})=0$,  $\forall \mathcal{L}$ non connected set such that $(\gamma\cup \delta)\subseteq \mathcal{L}$ when   $i'_{pa_D(T_h)} \in \mathcal{K}_{\gamma,\delta}^{pa_D(T_h)}$.\\
		\item[iii)]Each (CS3) holds iif the   regression parameters 
		in formula (\ref{eq:regression_param}) satisfy $\sum_{t\in C }\beta^{\gamma}_{t}(i'_{t})=0$, where $i'_t=\in \mathcal{K}^{pa_G(\gamma)}_{\gamma,\delta}\cap \mathcal{I}_t$ . \\
	\end{itemize}
	
\end{theorem}

\begin{example}
Let us consider the CGM in Figure \ref{nicolussi:fig1} (a) where there are two components. The first is composed of the purely dependent variables $3$, $5$ and $6$ while in the second there are the purely covariates $1$ and $2$. Then the following parameters fully describe the relationships among the 5 variables.
\small
\[
\begin{array}{lll}
\eta^{123}_{3}(i_3|i_{12})=&\beta_{\emptyset}^{3}+\beta_{1}(i_1)^{3},\; \quad &\forall i_3\in \mathcal{I}_3-1,\, \forall i_{12}\in \mathcal{I}_{12}\\
&\\
\eta^{123}_{4}(i_4|i_{12})=&\beta_{\emptyset}^{4}+\beta_{1}(i_1)^{4}+\beta_{2}^{4}(i_2)+\beta_{12}^{4}(i_{12}), &\forall i_4\in \mathcal{I}_4-1,\, \forall  i_{12}\in \mathcal{I}_{12} \\
&\\
\eta^{123}_{4}(i_4|i_{12})=&\beta_{\emptyset}^{4},\; &\forall  i_{12}\in \mathcal{I}_{12} \\
&\\
\eta^{123}_{35}(i_{35}|i_{12})=&\beta_{\emptyset}^{35}+\beta_{1}(i_1)^{35}+\beta_{2}^{5}(i_2)+\beta_{12}^{35}(i_{12}), &\forall i_{35}\in \mathcal{I}_{35}-1,\, \forall  i_{12}\in \mathcal{I}_{12} \\
&\\
\eta^{123}_{45}(i_{45}|i_{12})=&\beta_{\emptyset}^{45}+\beta_{1}(i_1)^{45}+\beta_{2}^{45}(i_2)+\beta_{12}^{45}(i_{12}), &\forall i_{45}\in \mathcal{I}_{45}-1,\, \forall  i_{12}\in \mathcal{I}_{12} \\

\end{array}
\]
\end{example}

\section{Application}
\label{sec_5}

In this section we study the relationships among a set of variables by using the regression model with CS independences as presented in Section \ref{sec_4}. At first we collect the variables in component according to their nature and the possible regression model that we want to study. \\ 
Several graphical models were tested and in each of them the likelihood ratio test $G^2$  is carried out. The $G^2$ compares the model under investigation   with the saturated (unconstrained) one; under the null hypothesis the $G^2$ follows the $\chi^2$ distribution with $df$ equal to the difference between the free parameters in the two models. We reject all models with a \textit{p-value} lower than $0.05$. Among the non rejected models, we choose the one with greatest  Akaike Information Criterion (AIC) and  Bayesian Information Criterion (BIC).\\
Since to testing all possible models, particularly when we handle with CS independencies, is computationally expensive, we implement a  three steps procedure to achieve the best SGCM of type IV. At first,  we carried out an exploratory phase where we test all CGM with only one missed arc in order to the have an overview of the weakest relationships. Then, we consider as \textit{reduced} model  the CGM without the arcs that have lead to a \textit{p-value} greater than $0.05$ in the previous step. Starting from the \textit{reduced} models we add one by one all removed arcs. We choose the CGM with greatest AIC and BIC. \\
A further simplification of the CGM is obtained evaluating the model with  the highest order parameters constrained to zero.\\ 
Finally, once obtained the best CGM we move on to further simplification by testing the CS independencies by simplifying the conditional ones that have lead to reject the model.
\subsection{Innovation Study Survey 2010-2012}
\begin{figure}
	\centering
	\includegraphics[width=0.4\linewidth]{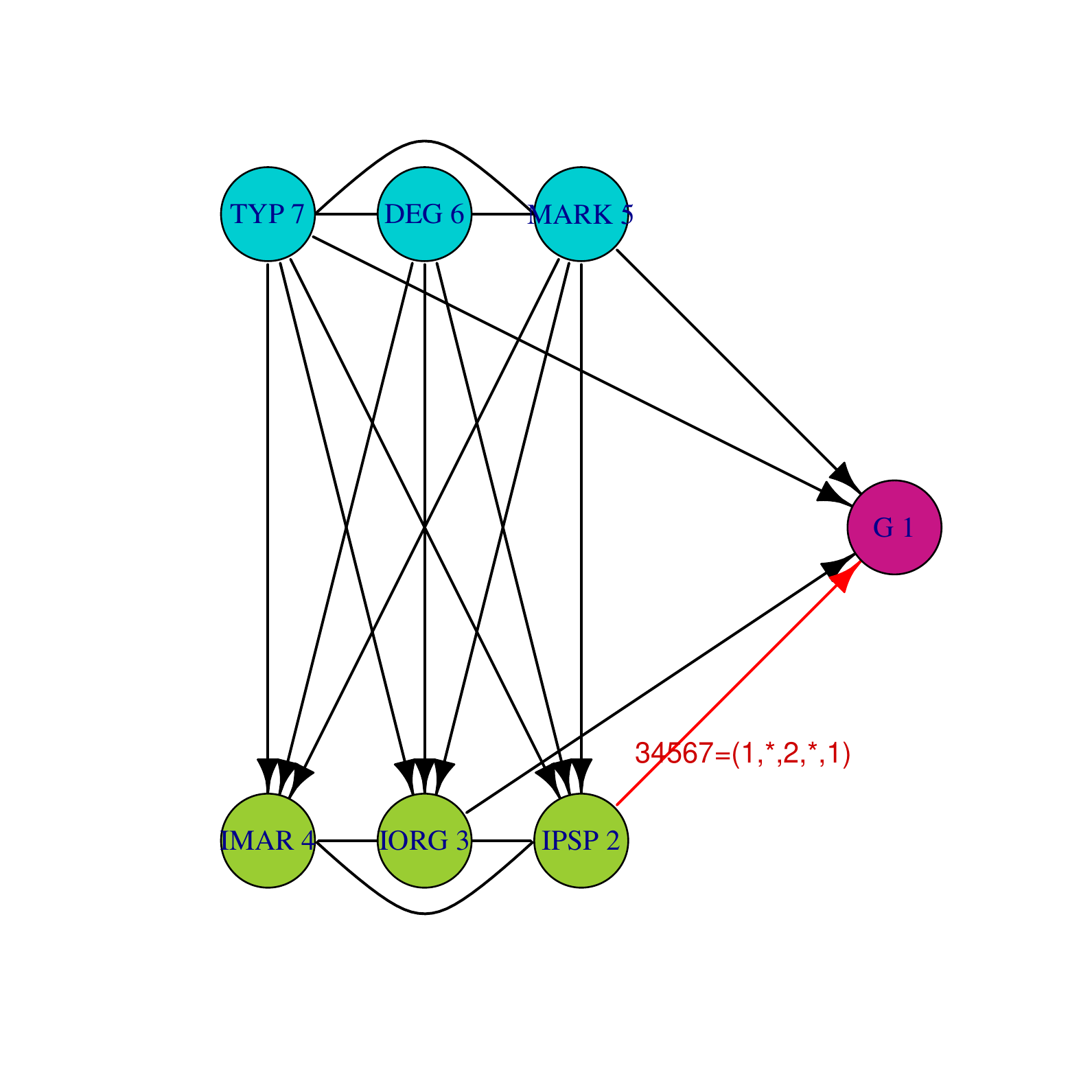}
	\caption{SCGM of type IV with components $T_1=(5,6,7)$, $T_2=(2,3,4)$ and $T_3=(1)$.}
	\label{fig.SGCM_IV}
\end{figure}

In this section we apply the proposed model on a real dataset.  
Our aim is to build a chain regression model that study the effect of the innovation in some aspects of the enterprise's life on the revenue growth without omitting the main features of the enterprise. Thus, we collect the following variables from the  survey on the innovation status of small and medium Italian enterprises during the $2010-2012$ \cite{ISTAT}. 
At first, as  pure response  we consider the \textit{revenue growth} variable  in 2012, \textbf{GROW} (Yes, No) henceforth denoted as variable \textbf{1}. Then, as mixed variables, we take into account the innovation through three dichotomous variables referring to the period 2009-2012: \textit{innovation in products or services or  production line or investment in R\&D}, \textbf{IPR} (Yes, No), \textit{innovation in organization system}, \textbf{IOR} (Yes, No) and \textit{innovation in marketing strategies}, \textbf{IMAR} (Yes, No), henceforth denoted as variables \textbf{2, 3} and \textbf{4}, respectively. Finally, the role of purely covariates is entrusted to variables concerning the firm's featuring in 2009-2012: the \textit{main market (in revenue terms)}, \textbf{MRKT} (A= Regional, B= National, C= International), the \textit{percentage of graduate employers}, \textbf{DEG} (1= $0\%\vdash10\%$, 2= $10\%\vdash50\%$, 3=$50\%\vdash100\%$) and the \textit{enterprise size}, \textbf{DIM} (1= Small, 2= Medium), henceforth denoted as variables \textbf{5, 6} and \textbf{7}, respectively. The survey covers $18697$ firms, collected in a $2\times2\times2\times2\times3\times3\times2$ contingency table. \\
In order to analyse this dataset, we build a chain graph with three components according to the nature of the variables, so in the first component we collect the firm's features variables (5,6,7), 
in the second component the innovations variables (2,3,4) 
and in the third component the revenue growth variabl (1).
\\
In the explanatory phases, we tested the independencies associated to all CGM of type IV with only one missed arc on the HMMM associated. Thus, according to the formula (\ref{marginal}), we considered the following marginal sets $\left\{(5,6,7);\,(2,5,6,7);\,(3,5,6,7);\,(4,5,6,7);\,(2,3,5,6,7);\,(2,4,5,6,7);\,\right.$ $\left.(3,4,5,6,7);\, (2,3,4,5,6,7);\,(1,2,3,4,5,6,7)\right\}$. The parameters associated to the dichotomous variables were based on the \textit{baseline} logits, while, the variables with three modalities have been coded with the  \textit{local} logits. We found the three eligible conditional independencies 
\begin{description}
\item[\textbf{(a)}] $1\perp 2|34567$,
\item[\textbf{(b)}]$1\perp 4|23567$,
\item[\textbf{(c)}] $1\perp 6|23457$.
\end{description}
 By testing the combination of these  independencies whom results are reported in Table \ref{tab_CGM}, we choose the HMMM characterized by the \textbf{(b)} and \textbf{(c)}, reported at the  third row in Table \ref{tab_CGM}, since it is the only model with a $p-value> 0.05$. \\
\begin{table}
	\centering
	\begin{tabular}{lrrrrr}
		\hline
		Independencies	& Gsq & df & pval & AIC & BIC \\ 
		\hline
		\textbf{(a)}, \textbf{(b)}& 139.74 & 108 & 0.02 & -220.26 & 1190.24 \\  
		\textbf{(a)}, \textbf{(c)}& 168.57 &120 &0.00 &-167.42 &1149.04\\
		\textbf{(b)}, \textbf{(c)} & 141.34 & 120 & \textbf{0.09} & -194.66 & 1121.81 \\ 	
		\textbf{(a)}, \textbf{(b)} ,\textbf{(c)}& 180.97 & 132 & 0.00 & -131.03 & 1091.40 \\ 
		\hline
	\end{tabular}
	\caption{HMMM which combining the three independencies \textbf{(a)} $1\perp 2|34567$, \textbf{(b)}$1\perp 4|23567$ and
		\textbf{(c)} $1\perp 6|23457$.}	\label{tab_CGM}
\end{table}
However, from the explanatory phases, there are some clues that independencies between variables $1$ and $2$ could be. Thus, among the \textbf{(b)} and \textbf{(c)}, we took into account also the independence \textbf{(a)} and we test all possible CS independencies originated from this last. The preferred model is described by the conditional independencies \textbf{(b)} and \textbf{(c)} and by the CS independence $1\perp2|34567=(1,*,3,*,1)$ that is when there are no innovation in  \textbf{IORG}, when the innovation \textbf{IMAR} assume any modality, when the firm works in an international market, when the percentage of degree employers is whatever and when the firm is small. In correspondence of this model we have \texttt{df=121}, \texttt{Gsq=141.83}, \texttt{p-val=0.09}, \texttt{AIC=-192.17}, \texttt{BIC=1116.46}.\\
The SCGM associated to this model is displayed in Figure \ref{fig.SGCM_IV}. Note that, in the stratum, the conditioning variables $3$ and $5$ are not set to a specific modality because they do not satisfy the condition for a CS independence as summarized in the Remark \ref{rem_stratum}. \\
By looking at the SCGM in Figure \ref{fig.SGCM_IV} as a regression chain model we can distinguish two regression structures, the first with the dependent variable \textbf{G} with all the others like covariates and the second with the innovation variables as dependent and the featuring variables as covariates.   The regression parameters of the two regression models are in Tables \ref{tab_1_par} and \ref{tab_2_par} respectively. In particular, in Table \ref{tab_1_par} we have the regression parameters of the only dependent variable $1$, thus the parameters are logarithms of logits concerning the variable $1$ (probability of no revenue growth against probability on revenue growth) evaluated in all the possible conditioning distributions. Generally speaking, when the parameters in Table \ref{tab_1_par} are less than zero it means that in the fitted probability having a revenue growth is greater than the probability to have not. In Table \ref{tab_1_par}, the conditional distributions where the difference between the two probabilities achieve high values (greater than 10 in absolute value) are highlight in bold. The cells where this disparity assume huge dimension in negative are where the conditioning distribution of the variables $3,4,5,6,7$ assume value equal to $(1,2,1,2,2,1)$ or $(1,1,1,2,2,1)$, on the other hand, the great disparity in positive is in the cells $(1,1,1,2,1,1)$ and $(1,1,1,1,1,1)$.
\begin{table}[ht]
	\centering
	\begin{tabular}{rr|rr|rr|rr}
		
		$i^*_{34567}$ &     $\eta^{\mathcal{M}}_{1}$ & $i^*_{34567}$ &     $\eta^{\mathcal{M}}_{1}$& $i^*_{34567}$ &      $\eta^{\mathcal{M}}_{1}$ & $i^*_{34567}$ &      $\eta^{\mathcal{M}}_{1}$ \\
		\hline
		222332 &    -0,4796 &     221222 &    -2,4468 &     222331 &    -0,3584 &     221221 &    -10,841 \\
		
		122332 &    -0,4483 &     121222 &   -11,5921 &     122331 &    -0,3141 &     \textbf{121221} &   \textbf{-32,7425} \\
		
		212332 &    -0,1707 &     211222 &     -2,071 &     212331 &     0,8602 &     211221 &    -4,9049 \\
		
		112332 &     0,6803 &     111222 &   -10,9506 &     112331 &     3,0365 &     \textbf{111221} &   \textbf{-27,4244} \\
		
		221332 &    -0,7988 &     222122 &     0,4845 &     221331 &    -1,8311 &     222121 &    -0,1683 \\
		
		121332 &     -2,102 &     122122 &     1,8001 &     121331 &    -3,9632 &     122121 &    -2,3828 \\
		
		211332 &    -0,5672 &     212122 &     0,3546 &     211331 &     0,1332 &     212121 &     0,7241 \\
		
		111332 &    -1,8448 &     112122 &      3,435 &     111331 &     0,6237 &     112121 &     0,2428 \\
		
		222232 &    -0,7417 &     221122 &    -1,4924 &     222231 &    -0,7225 &     221121 &    -9,1679 \\
		
		122232 &    -1,0322 &     121122 &    -2,4818 &     122231 &    -1,3533 &     \textbf{121121} &   \textbf{-16,5698} \\
		
		212232 &     0,8408 &     211122 &    -4,8401 &     212231 &     3,7374 &     \textbf{211121} &  \textbf{ -11,2031} \\
		
		112232 &     3,7735 &     111122 &    -5,8998 &     112231 &     9,8352 &     \textbf{111121} &    \textbf{-14,355} \\
		
		221232 &    -1,6234 &     222312 &    -0,4355 &     221231 &    -3,7989 &     222311 &    -0,3504 \\
		
		121232 &    -5,3967 &     122312 &    -0,6114 &     \textbf{121231} &    \textbf{-9,8567} &     122311 &    -0,2496 \\
		
		211232 &     0,6047 &     212312 &    -0,2184 &     211231 &     4,2275 &     212311 &     1,2512 \\
		
		111232 &    -0,7702 &     112312 &     0,5931 &     111231 &     7,1064 &     112311 &     5,1493 \\
		
		222132 &     0,0546 &     221312 &    -1,3541 &     222131 &     0,7491 &     221311 &    -3,6898 \\
		
		122132 &     1,3823 &     121312 &    -4,2052 &     122131 &     2,9369 &     121311 &    -7,1286 \\
		
		212132 &     1,0984 &     211312 &    -1,3974 &     212131 &     3,4766 &     211311 &    -0,3392 \\
		
		112132 &     4,7429 &     111312 &    -4,6389 &     \textbf{112131} &    \textbf{11,8076} &     111311 &     2,7044 \\
		
		221132 &    -0,8322 &     222212 &    -0,5193 &     221131 &    -2,8954 &     222211 &    -0,2307 \\
		
		121132 &    -0,1691 &     122212 &     -0,156 &     121131 &    -1,3416 &     122211 &     1,8533 \\
		
		211132 &    -0,5679 &     212212 &     1,1386 &     211131 &    -0,1813 &     212211 &      6,581 \\
		
		111132 &     2,3733 &     112212 &     5,9737 &    \textbf{ 111131} &    \textbf{12,1091} &    \textbf{ 112211} &    \textbf{22,5741} \\
		
		222322 &    -0,2048 &     221212 &    -2,3837 &     222321 &     -0,5180 &     221211 &    -6,0503 \\
		
		122322 &    -0,7378 &     121212 &    -6,7821 &     122321 &     -3,719 &     121211 &    -9,1221 \\
		
		212322 &    -0,4051 &     211212 &    -0,4052 &     212321 &    -0,0053 &     211211 &      7,674 \\
		
		112322 &    -0,5112 &     111212 &    -0,7885 &     112321 &    -4,0535 &     \textbf{111211} &    \textbf{29,7031} \\
		
		221322 &    -0,9409 &     222112 &     0,5425 &     221321 &    -4,3148 &     222111 &     1,0062 \\
		
		121322 &    -4,5919 &     122112 &     2,4386 &    \textbf{ 121321} &   \textbf{-13,8434} &     122111 &     4,9567 \\
		
		211322 &    -2,1517 &     212112 &     1,6583 &     211321 &    -4,0426 &     212111 &     4,3403 \\
		
		111322 &    -7,5744 &     112112 &      7,265 &     \textbf{111321} &   \textbf{-16,6193} &    \textbf{ 112111} &    \textbf{20,1193} \\
		
		222222 &    -0,4359 &     221112 &    -1,6496 &     222221 &    -2,1184 &     221111 &    -6,8449 \\
		
		122222 &    -2,3083 &     121112 &    -1,1323 &    \textbf{ 122221} &    \textbf{-10,308} &     121111 &    -2,5094 \\
		
		212222 &     0,6317 &     211112 &    -2,2924 &     212221 &     1,4343 &     211111 &    -3,1236 \\
		
		112222 &     2,0696 &     111112 &     2,6188 &     112221 &    -5,3214 &    \textbf{ 111111} &   \textbf{ 26,1675} \\
		
	\end{tabular}  
	\caption{Regression parameters concerning the dependent variable 1 with covariates $2,3,4,5,6,7$. The $i^*_{34567}$ in table refers to the conditional distribution  where the parameters concerning the dependent variable $1$ is evaluated as in formula (\ref{eq:regression_param}), i.e. $\eta^{\mathcal{V}}_{1}(i_1=2|i^*_{234567})$ where $V=123456$. } 
	\label{tab_1_par}
\end{table}

In Table \ref{tab_2_par} we report the regression parameters concerning the combination of the three dependent variables $2$, $3$ and $4$. In particular, from the 2th to the 4th column there are  the parameters associated to the single variables, thus the parameters are logarithms of logits. In the columns 5 to 7 there are the logarithms of contrasts of logists and in the last column there are the third order parameters associated to the variables $2,3,4$. In the first group of columns there is a prevalence of positive parameters which highlight a trend where the probability to make any innovation is lower than the probability to do not, wherever  the conditioning distribution is. In the column  5 to 7 there are the pairwise comparison between the different kinds of innovation. Where the parameters are negative, such as in the 4th and in the 6th columns, the probability of concordance between the two innovations considered (i.e. innovation in both aspects or no innovation in both aspects) is lower than the probability of discordance. The opposite case occurs in the 5th column. 

\begin{table}[ht]
	\centering
	
	\begin{tabular}{r|rrrrrrr}
		
		$i^{*}_{567}$ &     $\eta_{2}^{2567}$ &     $\eta_{3}^{3567}$ &     $\eta_{4}^{4567}$ &      $\eta_{23}^{23567}$ &     $\eta_{24}^{24567}$ &    $\eta_{34}^{34567}$ &   $\eta_{234}^{234567}$ \\
		\hline
		332 &     0,6831 &     0,5360 &    -0,1749 &    -1,7706 &    -1,9050 &    -1,5165 &     0,9079 \\
		
		232 &     0,6275 &     0,5631 &    -0,0295 &    -1,8384 &    -1,3781 &    -1,7466 &     0,6467 \\
		
		132 &     0,2105 &     0,5281 &    -0,0161 &    -1,8344 &    -1,0737 &    -1,7653 &     0,4587 \\
		
		322 &     1,2485 &     0,6050 &    -0,0662 &    -1,9450 &    -1,2532 &    -1,5024 &    -1,1644 \\
		
		222 &     1,6045 &     0,6567 &     0,0426 &    \textbf{-2,0638} &     0,3596 &    -2,0162 &    -2,8816 \\
		
		122 &     1,3508 &     0,7910 &     0,3162 &    -1,9630 &     0,6518 &    -1,5455 &    -3,1860 \\
		
		312 &     0,7214 &    -0,0583 &    -0,4054 &    -1,8619 &    -1,8553 &    -1,5649 &    -0,5257 \\
		
		212 &     1,3267 &     0,3062 &     0,1200 &    -1,9385 &    -0,8871 &    -1,5361 &    -1,4836 \\
		
		112 &     1,0679 &     0,3998 &     0,4696 &    -1,8852 &    -0,2482 &    -1,5967 &    -1,4493 \\
		
		331 &     0,3001 &     0,1707 &    -0,0851 &    -1,3967 &    -0,9175 &    -1,7610 &    -1,6513 \\
		
		231 &     0,2883 &     0,3485 &     0,5260 &    -1,7401 &     0,8988 &    -2,2774 &    -3,8131 \\
		
		131 &     0,3271 &     0,7048 &     0,7575 &    -0,8155 &     1,2579 &    -2,0448 &    -4,1997 \\
		
		321 &     1,5258 &     0,7076 &     0,2295 &    -1,5524 &     0,7126 &    -2,2999 &    -7,6228 \\
		
		221 &     2,1658 &     1,0819 &     1,0864 &    \textbf{-2,2763} &     \textbf{4,9992} &    \textbf{-3,9665} &   \textbf{-15,2039} \\
		
		121 &     \textbf{2,8376 }&     \textbf{1,8068 }&     1,4223 &    -0,7067 &     \textbf{5,1590} &    \textbf{-2,7446} &   \textbf{-15,6388} \\
		
		311 &     1,0988 &     0,0431 &     0,0358 &    -1,4920 &     0,2885 &    -2,0071 &    -5,4161 \\
		
		211 &    2,6186 &     1,1211 &     1,1570 &    -2,0384 &     3,9245 &    -2,2946 &   -10,4366 \\
		
		111 &     \textbf{2,8229} &     \textbf{1,7590} &     \textbf{2,2578} &    -0,6703 &     4,7713 &    -2,1910 &  -10,4318 \\

	\end{tabular}  
	\caption{Regression parameters concerning the dependent variables $2,3,4$  with covariates $5,6,7$.The $i^*_{567}$ in table refers to the conditional distribution  where the parameters concerning the dependent variables $2,3,4$ is evaluated as in formula (\ref{eq:regression_param}), i.e. $\eta^{\mathcal{M}}_{1}(i_1=2|i^*_{567})$. } 
	\label{tab_2_par}
\end{table}

In conclusion, the output of this application shows a little aspects of the things that we can derive from the application of this models. For instance, once fitted the model it can be used to forecast the values of some dependent variables given the covariate, or again, looking the regression parameters it is possible to define a strategy where to invest. The possibility are several, it depends on the aim of the analysis, the HMMM parameters (that are not listed here) can be used to study the relationships among the variables.
\section{Conclusions}
\label{sec_6}
In this work we provide several results in the environment of context-specific independences. At first, we focus on the problem to handle with ordinal variables where it is more useful to use parameters based on the \textit{local} or \textit{continuation} logits compares to  the classical ones based on the \textit{baseline} logits. In this case, not only we confirm the results on \textit{baseline} logits such as provided in \cite{nyman2016}, even if in the marginal models, but we provide the results in the case of \textit{local} and \textit{continuation} parameters. \\
Further, we focus on the problem of the graphical representation. We take advantage from the well known relationships between the HMMM and the chain graphical models, in particular the type IV, and we extend the so-called stratified graph to this case.\\
Finally, we provide the advantage of the use of CS independencies in the Chain Regression models, \cite{marchetti2011} where the CS independencies can simplify the models.\\
The application shows a small part of the potentiality of this work.

\appendix
\section*{Appendices}
\addcontentsline{toc}{section}{Appendices}
\renewcommand{\thesubsection}{\Alph{subsection}}
\subsection{Proofs and further results}
\label{appendix_1}    
\begin{lemma}                                                 
	\label{lemma1} 
	Given a HMM parameter $\eta_{\mathcal{L}C}^{\mathcal{M}}(i_{\mathcal{L}C})$, where the set $\mathcal{L}C$ is the union of two sets of variables belonging in $\mathcal{M}$, it can be decomposed as follow                 
	\begin{equation}                                                                                                            
	\eta_{\mathcal{L} C}^{\mathcal{M}}(i_{\mathcal{L} C})=\sum_{\substack{J\subseteq C\\ J\neq \emptyset}}(-1)^{|J|+1} \eta_{\mathcal{L} (C\backslash J)}^{\mathcal{M}}(i_{\mathcal{L}(C\backslash J)}|i^{*}_{J})+(-1)^{|C|}\eta_{\mathcal{L} }^{\mathcal{M}}(i_{\mathcal{L}}|i_{C})
	\label{nic_fact}
	\end{equation}
\end{lemma}
\paragraph*{Proof of Lemma \ref{lemma1}}
From Proposition (1) of \cite{bartolucci2007} any parameter $\eta_{\mathcal{L}C}^{\mathcal{M}}(i_{\mathcal{L}C})$ can be rewritten as
\begin{equation}
\eta_{\mathcal{L}C}^{\mathcal{M}}(i_{\mathcal{L}C})=\sum_{J\subseteq C}(-1)^{|C\backslash J|}\eta^{\mathcal{M}}_{\mathcal{L}}(i_{\mathcal{L}}|i_{C\backslash J}i^*_{J})
\label{pr1}
\end{equation}
where $\eta^{\mathcal{M}}_{\mathcal{L}}(i_{\mathcal{L}}|i_{C\backslash J}i^*_J)$ is the HMM parameter $\eta^{\mathcal{M}}_{\mathcal{L}}$ evaluated in the conditional distribution where the variables in $C\backslash J$ assume values $i_{C\backslash J}$ and the variables in $J$ are set to a reference modality $i^*_J$. \\ 
When the set $C$ is only one variable, $C=\gamma_1$, the decomposition in formula (\ref{pr1}) becomes
\begin{equation}
\eta^{\mathcal{M}}_{\mathcal{L}C}(i_{\mathcal{L}C})= \eta^{\mathcal{M}}_{\mathcal{L}}(i_{\mathcal{L}}|i^*_{\gamma_1})-\eta^{\mathcal{M}}_{\mathcal{L}}(i_{\mathcal{L}}|i_{\gamma_1})
\label{l1_c1}
\end{equation}
that corresponds to  formula (\ref{nic_fact}).\\
When two variables belong to the  set $C$, $C=\left\{\gamma_1,\gamma_2\right\}$, by applying the formula  (\ref{pr1}) only to $\gamma_{1}$ we get
\begin{equation}
\eta^{\mathcal{M}}_{\mathcal{L}C}(i_{\mathcal{L}C})= \eta^{\mathcal{M}}_{\mathcal{L}\gamma_2}(i_{\mathcal{L}\gamma_2}|i^*_{\gamma_1})-\eta^{\mathcal{M}}_{\mathcal{L}\gamma_2}(i_{\mathcal{L}\gamma_2}|i_{\gamma_1});
\label{l1_c12}
\end{equation}
the second addend on the right hand side, can be further decomposed by using the (\ref{pr1}) as:
\begin{equation}
\eta^{\mathcal{M}}_{\mathcal{L}\gamma_2}(i_{\mathcal{L}\gamma_2}|i_{\gamma_1})= \eta^{\mathcal{M}}_{\mathcal{L}}(i_{\mathcal{L}}|i_{\gamma_1}i^*_{\gamma_2})-\eta^{\mathcal{M}}_{\mathcal{L}}(i_{\mathcal{L}}|i_{\gamma_1\gamma_2}).
\label{l1_c2_2}
\end{equation}
Now, by considering the HMM parameter $\eta^{\mathcal{M}}_{\mathcal{L}\gamma_1}(i_{\mathcal{L}\gamma_1}|i^*_{\gamma_2})$ and by applying the formula (\ref{pr1}), we get
\begin{equation}
\eta^{\mathcal{M}}_{\mathcal{L}\gamma_1}(i_{\mathcal{L}\gamma_1}|i^*_{\gamma_2})=\eta^{\mathcal{M}}_{\mathcal{L}}(i_{\mathcal{L}}|i^*_{\gamma_1\gamma_2})- \eta^{\mathcal{M}}_{\mathcal{L}}(i_{\mathcal{L}}|i_{\gamma_1}i^*_{\gamma_2}).
\label{l1_eq}
\end{equation}
Note that the  last addend on the right hand side of the (\ref{l1_eq}) is exactly the first addend on the right hand side of (\ref{l1_c2_2}). Thus, by replacing the (\ref{l1_c2_2}) and (\ref{l1_eq}) in the (\ref{l1_c12}) we get:
\begin{equation}
\eta^{\mathcal{M}}_{\mathcal{L}C}(i_{\mathcal{L}C})= \eta^{\mathcal{M}}_{\mathcal{L}\gamma_2}(i_{\mathcal{L}\gamma_2}|i^*_{\gamma_1})-\eta^{\mathcal{M}}_{\mathcal{L}}(i_{\mathcal{L}}|i^*_{\gamma_1\gamma_2})+
\eta^{\mathcal{M}}_{\mathcal{L}\gamma_1}(i_{\mathcal{L}\gamma_1}|i^*_{\gamma_2})
+
\eta^{\mathcal{M}}_{\mathcal{L}}(i_{\mathcal{L}}|i_{\gamma_1\gamma_2})
\label{l1_c2}
\end{equation}
that again corresponds to  formula (\ref{nic_fact}).\\
In general, when the set $C$ is composed of $k$ variables, $C=\left\{\gamma_1,\dots,\gamma_k\right\}$, we apply formula (\ref{pr1}) recursively, focusing on only one variable of $C$ each time, to any parameter in the formula without any index $i^*$ in the conditioning set. 
\begin{equation}
\eta^{\mathcal{M}}_{\mathcal{L}C}(i_{\mathcal{L}C})= 
\sum_{j=1}^{k}(-1)^{j+1}
\eta^{\mathcal{M}}_{\mathcal{L} C\backslash (\gamma_{jp} \gamma_j)}(i_{\mathcal{L}C\backslash(\gamma_{jp} \gamma_j)}|i^*_{\gamma_j}i_{\gamma_{jp}})+ (-1)^{|C|}\eta_{\mathcal{L}}^{\mathcal{M}}(i_{\mathcal{L}}|i_{C}).
\label{l1_c1k}
\end{equation}
where $\gamma_{jp}=\sum_{i=1}^{j-1}\gamma_{i}$.\\	
Now, we take into account all the parameters having both $i$ and $i^*$ in the conditioning set. Let us denote it as $\eta^{\mathcal{M}}_{\mathcal{L}}(i_{\mathcal{L}}|i_{A}i^*_{B})$. We can recognise this term in the last term of the right hand side of the decomposition \ref{l1_ck_2} obtained applying the  \ref{pr1} to $\eta^{\mathcal{M}}_{\mathcal{L} A}(i_{\mathcal{L} A}|i^*_{B})$:

\begin{equation}
\eta^{\mathcal{M}}_{\mathcal{L} A}(i_{\mathcal{L} A}|i^*_{B})= \sum_{\substack{J\subseteq A\\ J\neq \emptyset}} (-1)^{|A\backslash J|}\eta^{\mathcal{M}}_{\mathcal{L}}(i_{\mathcal{L}}|i^*_{B J}i_{A\backslash J})+\eta^{\mathcal{M}}_{\mathcal{L}}(i_{\mathcal{L}}|i^*_{B}i_{A})
\label{l1_ck_2}
\end{equation}
By replacing in formula (\ref{l1_c1k}) each addend like $\eta^{\mathcal{M}}_{\mathcal{L}}(i_{\mathcal{L}}|i_{A}i^*_{B})$ with the expression learned from formula (\ref{l1_ck_2}), and applying this procedure recursively to any  addend like $\eta^{\mathcal{M}}_{\mathcal{L}}(i_{\mathcal{L}}|i_{A}i^*_{B})$, we finally obtain  exactly the formula \ref{nic_fact}.

\begin{corol}
	\label{cor1_APP}
	A parameter $\eta_{\mathcal{L}}^{\mathcal{M}}$  can be decomposed as the sum of greater order parameters as follows:
		\begin{equation}
	\eta_{\mathcal{L}}^{\mathcal{M}}(i_{\mathcal{L}}|i_{C})=\sum_{J\subseteq C} (-1)^{|C\backslash J|}\eta_{\mathcal{L} J}^{\mathcal{M}}(i_{\mathcal{L} J}|i^{*}_{C\backslash J})
	\label{dec_2}
	\end{equation}
\end{corol}
\paragraph*{Proof of Corollary \ref{cor1_APP}}
From formula (\ref{nic_fact}), we isolate the last right term having
\begin{equation}
\begin{array}{ll}
(-1)^{|C|}\eta_{\mathcal{L}}^{\mathcal{M}}(i_{\mathcal{L}}|i_{C})&=\sum_{J\subseteq C} (-1)^{|J|}\eta_{\mathcal{L} (C\backslash J)}^{\mathcal{M}}(i_{\mathcal{L} (C\backslash J)}|i^{*}_{ J})\\
\\
\eta_{\mathcal{L}}^{\mathcal{M}}(i_{\mathcal{L}}|i_{C})&=\sum_{J\subseteq C} (-1)^{|C\backslash J|}\eta_{\mathcal{L}(C\backslash J)}^{\mathcal{M}}(i_{\mathcal{L} (C\backslash J)}|i^{*}_{ J})
\end{array}
\label{dec_3}
\end{equation}	
By replacing $C\backslash J$ with $J$ in the left side, we get exactly the decomposition in formula (\ref{dec_2}).

\paragraph*{Proof of Theroem \ref{T1_baseline}}
When the CS independence in formula (\ref{nic1}) holds, let us consider the parameters $\eta^{\mathcal{M}}_{\mathcal{L}}$ when $\mathcal{L}=A\cup B\cup C\subseteq \mathcal{M}$. From Lemma \ref{lemma1} we can decompose it as

\begin{equation}
\eta_{ABC}^{\mathcal{M}}(i_{ABC})=
\sum_{\substack{J\subseteq C\\ J\neq \emptyset}}(-1)^{|J+1|} \eta_{AB(C\backslash J)}^{\mathcal{M}}(i_{AB(C\backslash J)}|i^*_{J})+(-1)^{|C|}\eta_{AB}^{\mathcal{M}}(i_{AB}|i_{C})
\label{t1_dec}
\end{equation}
where $\eta^{ABC}_{AB}(i_{AB}|i_{C})$ is the marginal parameter $\eta^{ABC}_{AB}$ evaluated in the conditional distribution $(AB|C=i_C)$. The term $\eta^{ABC}_{AB}(i_{AB}|i_{C})$ is equal to zero if and only if the CS independence in formula (\ref{nic1}) holds. 
Thus, 
\begin{equation}
\begin{array}{ll}
\eta_{ABC}^{ABC}(i_{ABC})- \sum_{\substack{J\subseteq C\\ J\neq \emptyset}}(-1)^{|J+1|} \eta_{AB(C\backslash J)}^{ABC}(i_{AB(C\backslash J)}|i^*_{J})&=0\\
&\\
\eta_{ABC}^{ABC}(i_{ABC})+
\sum_{\substack{J\subseteq C\\ J\neq \emptyset}}(-1)^{|J|} \eta_{AB(C\backslash J)}^{ABC}(i_{AB(C\backslash J)}|i^*_{J})&=0\\
&\\
\sum_{J\subseteq C}(-1)^{|J|} \eta_{AB(C\backslash J)}^{ABC}(i_{AB(C\backslash J)}|i^*_{J})&=0\\
&\\
\sum_{c\in \mathcal{P}(C)}(-1)^{|C\backslash c|} \eta_{ABc}^{ABC}(i_{ABc}|i^*_{C\backslash c})&=0\\
\end{array}
\label{t1_b_1}
\end{equation}
Note that in the case of \textit{baseline} coding, the cell $i^*_{C\backslash c}$ is equivalent to $i^{**}_{C\backslash c}$ thus  the parameters $\eta_{ABc}^{ABC}(i_{ABc}|i^*_{C\backslash c})$ is  $\eta_{ABc}^{ABC}(i_{ABc})$.\\
Finally, by considering that the previous decomposition holds for each set $v\in\mathcal{V}=\left\{\left(\mathcal{P}(A)\setminus\emptyset\right)  \cup \left(  \mathcal{P}(B)\setminus \emptyset\right) \right\}$,  the formula (\ref{eq.teo1}) comes.

\paragraph*{Proof of Theorem \ref{T2_local}}
From the proof of Theorem \ref{T1_baseline}, the decomposition in formula (\ref{t1_b_1}) still holds. However, by using the \textit{local} logits $i^*_{C\backslash c}\neq i^{**}_{C\backslash c}$ and the identity $\eta^{ABC}_{ABc}(i_{ABc}|i^*_{C\backslash c})=\eta^{ABC}_{ABc}(i_{ABc})$ does not hold any more because in \textit{local} logits $i^*_{C\backslash c}$ is equal to $\cap_{j\in C\backslash c} (i_j+1)$  while the parameter $\eta^{ABC}_{AB}(i_{AB})$ is built in the conditional distribution where the variables in $C$ assume the reference value $I_C$. 
Note that  $\eta^{ABC}_{ABc}(i_{ABc}|i_{C\backslash c}+1)$ does not belong to this parametrization. 
Now we remark that  between the \textit{baseline} parameters, $\eta_{\textit{b}}$, and the \textit{local} parameters $\eta_{\textit{l}}$, the following relationship holds:
\begin{equation}
\eta_{\textit{b}\mathcal{L}}^{\mathcal{M}}(i_{\mathcal{L}})=\sum_{i'_{\mathcal{L}}\geq i_{\mathcal{L}}}\eta_{\textit{l}\mathcal{L}}^{\mathcal{M}}(i'_{\mathcal{L}}).
\label{relation_b_l}
\end{equation}
When the variables in the conditioning set $C$ are based on  \textit{local} logits, it is enough to apply the decomposition in  (\ref{relation_b_l}) only on the variables in the parameter $\eta_{ABC}^{ABC}(i_{ABC})$ in order to have a \textit{baseline} approach in $C$. Thus we can rewrite  (\ref{t1_b_1}) as:
\begin{equation}
\sum_{c \in \mathcal{P}(C)}(-1)^{|C\backslash c|}\sum_{i'_c \geq i_c}\eta_{ABc}^{ABC}(i_{AB}i'_c|I_{C\backslash c})=0	
\label{eq_t2_rel}
\end{equation}
where $\eta_{ABC}^{ABC}$ are the \textit{local} parameters and are exactly the same of formula (\ref{nic2}).
As in proof of the Theorem \ref{T1_baseline}, the previous equivalence must hold for each subset $v$ of $A\cup B$ with at least one element in $A$ and one element in $B$.

\paragraph*{Proof of Corollary \ref{cor1}}
When $i_C$ in $\mathcal{K}$ is equal to the last modalities $I_C$, for each $c\subseteq C$ and $c\neq \emptyset$, the parameters $\eta^{\mathcal{M}}_{vC}(i_{vC})=0$ by definition, thus formula (\ref{nic2}) in Theorem \ref{T2_local} becomes $\eta^{\mathcal{M}}_{v}=0$ $\forall v\in \mathcal{V}$.  
When $\mathcal{K}=\left\{ (I_{C\backslash j})\cap(I_{j}-1)\right\}$, that is the modality of every variable is equal to the last but the variable $j$ assumes the modality $I_j-1$, the constraints become $\eta^{\mathcal{M}}_{v}(i_v)=0$  and $\eta^{\mathcal{M}}_{vj}(i_{vj})=0$. Applying this procedure recursively for each $i'_C$ we obtain the constraints in formula (\ref{eq_cor1}).

\paragraph*{Proof of Theorem \ref{teo3}} 
For a $c\in \mathcal{P}(C)$ and a $v\in \mathcal{V}$ we consider the parameters $\eta_{vc}^{\mathcal{M}}(i_{vc})$. Note that, each variable $X_j$ in $C$ assumes value in $i_j$ or in $((i_j+1)+\dots+I_j)$ when it drop in the reference modality. But since in each of these distributions the CS independence (\ref{nic_cs_1}) holds the parameters are equal to zero.

\paragraph*{Proof of Theorem \ref{teo:param}}
By applying formula (\ref{dec_2}) (Corollary \ref{cor1_APP})  to the HMM parameters in formula (\ref{eq:regression_param})  evaluated on the conditional distribution $i_A|i_{pa_D(T_h)}$ we obtain: 
\begin{equation}
\eta^{A\cup pa_D(T_h)}_{A}(i_A|i_{pa_D(T_h)})=\sum_{t\subseteq pa_D(T_h)}(-1)^{|t|}\eta^{A\cup pa_D(T_h)}_{tA}(i_{tA}|i^{*}_{pa_D(T_h)\backslash t}).
\label{eq_theo_4.1}
\end{equation}
Thus, the $\beta$ parameters correspond to the addends on the left side of (\ref{mix}).

\paragraph*{Proof of Theorem \ref{parametrizz}}
When $i_{pa_D(T_h)}= I_{pa_D(T_h)}$, each variable in $pa_D(T_h)$ assumes the last modality, the formula (\ref{eq:regression_param}) becomes
\begin{equation}
\eta^{A\cup pa_D(T_h)}_{A}(i_A|I_{pa_D(T_h)})=\beta_{\emptyset}^{A}=\eta^{A\cup pa_D(T_h)}_{A}(i_A).
\label{eq.teo.4.2}
\end{equation}
Now, by considering only one variable $X_j\in pa_D(T_h)$ with $i_j\neq I_j$ and the remaining $pa_D(T_h)\backslash X_j$ setting equal to $I_{pa_D(T_h)\backslash X_j}$, we have
\begin{equation}
\eta^{A\cup pa_D(T_h)}_{A}(i_A|i_{pa_D(T_h)})=\beta_{\emptyset}^{A}+\beta_{j}^{A}(i_j)=\eta^{A\cup pa_D(T_h)}_{A}(i_A)-\eta^{A\cup pa_D(T_h)}_{Aj}(i_{Aj})
\label{eq.teo.4.3}
\end{equation}
where we can isolate the term $\eta^{A\cup pa_D(T_h)}_{Aj}(i_{Aj})$. In analogous way we can obtain all HMM parameters  defined on $T_h\cup pa_D(T_h)$ for each $h=1,\dots,s$. 
The other parameters are exactly listed in formula (\ref{mix}).
\paragraph*{Proof of Theorem \ref{regression_constraints}}
Let us remember that, given an independence \added[id=Fe]{like} $A\perp B|C$, the probability distribution of $ABC$ obeys to the independence if and only if the HMM parameters $\eta_{abc}^{\mathcal{M}}=0$ where $a\subseteq A$, $b\subseteq B$, $c\subseteq C$ and   $a,b\neq\emptyset$.\\
Point \textit{i)}. All the HMM parameters in formula (\ref{mix}) refer to the non connected sets of variables involved in formula \textbf{(C1)} of the Markov Properties in (\ref{SMP_IV}), thus they are equal to zero.\\
Point \textit{ii)}. In this case the variables $\gamma$ and $\delta$ belong both to the same component $T_h$. Let us focusing on the parameters ground on the \textit{baseline} logits.  Thus, by representing the constraints in  \textbf{(CS2)}   in formula  (\ref{SMP_IV}), through the constraints in formula  (\ref{eq.teo1}), we get  becomes $\sum_{t\subseteq pa_D(T_h)}(-1)^{|pa_D(T_h)\backslash t|}\eta_{\gamma\delta t}^{\mathcal{M}}(i_{\gamma\delta }i'_t)$ where the marginal distribution is $\mathcal{M}=\gamma\cup \delta \cup pa_D(T_h)$,  $i_{\gamma\delta }=i_\gamma\cap i_\delta$ and $i'_t\in \mathcal{K}_{\gamma,\delta}^{pa_D(T_h)}\cap\mathcal{I}_{t}$. By using the equivalence in formula (\ref{formula}) and remembering that in the case of \textit{baseline} logits the index $i^*$ is equal to the index $i^{**}$, we obtain $\sum_{t\in \subseteq pa_D(T_h)}\beta^{\gamma\delta}_{t}(i'_t)=0$. Thus, according to  formula (\ref{eq:regression_param}),  we have $\eta_{\gamma \delta}^\mathcal{M}(i_{\gamma\delta}|i'_{pa_D(T_h)})=0$, $\forall i_{pa_D(T_h)}\in \mathcal{K}_{\gamma,\delta}^{pa_D(T_h)}$. \\
On the other hand, by considering the parameters based on the \textit{local} logits, we use the constraints in formula (\ref{nic2}) that with \textbf{(CS2)} becomes $\sum_{t\subseteq pa_D(T_h)}(-1)^{|pa_D(T_h)\backslash t|}\sum_{i_t\geq i'_t}\eta_{\gamma\delta t}^{\mathcal{M}}(i_{\gamma\delta}i_t)$ for $i'_t\in \mathcal{K}_{\gamma,\delta}^{pa_D(T_h)}\cap\mathcal{I}_{t}$ where $\mathcal{M}=\gamma \cup \delta\cup pa_D(T_h)$, as above. But, according to the relationship between \textit{local}  and \textit{baseline} parameters in (\ref{relation_b_l}), we can rewrite the previous constrast as $\sum_{t\subseteq pa_D(T_h)}(-1)^{|pa_D(T_h)\backslash t|}\eta_{\gamma\delta t}^{\mathcal{M}}(i_{\gamma\delta t}|i^{*}_{pa_D(T_h)\backslash (\gamma \delta t)})$ for $i_t\in \mathcal{K}\cap\mathcal{I}_{t}$. Thus the proof follows the \textit{baseline} case.
\\
Point \textit{iii)}. The \textbf{(CS3)} in formula  (\ref{SMP_IV}) concerns the independencies with $\gamma\in T_h$ and $\delta \in pa_D(T_h)\backslash pa_G(\gamma)$. Let us focusing on the parameters based on \textit{baseline} logits.  The constraints on parameters in formula  (\ref{eq.teo1}) can be express as $\sum _{t\subseteq P}(-1)^{|P\backslash t|}\eta^{\mathcal{M}}_{\gamma t}(i_{\gamma }i'_t|i^*_{P\backslash t})=0$,  where $P=pa_D(T_h)\backslash pa_G(\gamma)$, $\mathcal{M}=\gamma\cup pa_D(T_h)$,  $\forall i_{\gamma}\in \mathcal{I}_{\gamma}$ and $ i'_{t}\in \mathcal{K}_{\gamma,\delta}^{pa_G(\gamma)}$. Now, considering the relationship between HMM and regression parameters express in 
formula (\ref{formula}), we get  $\sum _{t\in P}\beta^{\gamma}_{t}(i'_t)=0$ $\forall i'_t\in \mathcal{K}_{\gamma,\delta}^{pa_G(\gamma)}\cap\mathcal{I}_{t}$. 
In the case of \textit{local} coding we must remember that  $\eta^{\gamma T}_{\gamma t}(i_{\gamma t}|i^{**}_{T\backslash t})=\sum_{i'_t\geq i_t} \eta^{\gamma T}_{\gamma t}(i_{\gamma}i'_t|i^*_{T\backslash t})$ as showed in formula (\ref{relation_b_l}). Thus the reasoning  done for \textit{baseline} can be generalized to the parameters based on the  \textit{local} logits.

\bibliography{References}
\bibliographystyle{authordate3} 
\newpage
\listofchanges[style=<list|summary>]	
\end{document}